# Forethought

# Preparing for the Intelligence Explosion

Will MacAskill & Fin Moorhouse

March 2025

# Contents

# Preparing for the Intelligence Explosion

# Abstract

AI that can accelerate research could drive a century of technological progress over just a few years. During such a period, new technological or political developments will raise consequential and hard-to-reverse decisions, in rapid succession. We call these developments *grand challenges*.

These challenges include new weapons of mass destruction, AI-enabled autocracies, races to grab offworld resources, and digital beings worthy of moral consideration, as well as opportunities to dramatically improve quality of life and collective decision-making.

We argue that these challenges cannot always be delegated to future AI systems, and suggest things we can do today to meaningfully improve our prospects. AGI preparedness is therefore not just about ensuring that advanced AI systems are aligned: we should be preparing, now, for the disorienting range of developments an intelligence explosion would bring.

# 1. Introduction

There's now a serious chance we will see AI far smarter than humans within the coming decade.[1] The best language models in 2019 were barely coherent;[2] by early 2023 they could answer questions fluently and with more general knowledge than any living person. The best models in early 2023 performed little better than random guessing on reasoning-intensive science questions; today they outscore Ph.D-level experts.[3]

But there is a common view among those who expect superintelligent AI that outcomes are effectively all-or-nothing,[4] and depend on one challenge: AI alignment.[5] Either we fail to align AI, in which case humanity is permanently disempowered; or we succeed at aligning AI, and then we use it to solve all of our other problems.

---

1. The forecasting platform Metaculus features a question titled, ['When will the first general AI system be devised, tested, and publicly announced?'](), operationalised across a range of cognitive and robotics tests, including the ability to "reliably pass a 2-hour, adversarial Turing test during which the participants can send text, images, and audio files (as is done in ordinary text messaging applications) during the course of their conversation.". As of [February 5, 2025](), the community prediction (a recency-weighted median of all forecasts) was **mid-2030**.

2. AI Digest, ['How Fast Is AI Improving?']()

3. Epoch AI, ['AI Benchmarking Dashboard'](). This is the "GPQA" benchmark. "Ph.D-level experts" here refers to people who are enrolled in or have completed a PhD program in biology, physics and chemistry. Questions are limited to these fields, and were designed by 61 Ph.D-level experts hired from an online freelancing platform. Upwards of 74% of questions have non-controversially correct answers. Correct answers were determined by agreement with other experts.

4. Yudkowsky, ['Complex Value Systems Are Required to Realize Valuable Futures']()., Barnes, ['Considerations on Interaction between AI and Expected Value of the Future']().

5. For more discussion of AI alignment, see [Ngo et al. (2021)]() and [Carlsmith (2024](), [2025]()).

6. By "AGI", we intend to follow the definition of "Expert AGI" given in Morris et al., ['Levels of AGI for Operationalizing Progress on the Path to AGI'](), which is performance exceeding "at least 90th percentile of skilled adults" across a "wide range of non-physical tasks, including metacognitive tasks like learning new skills".

This paper argues against this "all or nothing" view. Instead, we argue that AGI[6] preparedness should cover a wider range[7] of opportunities and challenges.

The development of AI that can meaningfully substitute for human research labour would probably drive very rapid technological development, compressing decades of scientific, technological and intellectual development into years or months. This will of course bring the potential for enormously improving quality of life, just as technological advances have regularly improved quality of life in the past. It could mean medical advances which drastically extend our healthy lifespans, enough material abundance to give everyone a lifestyle that today's billionaires would envy, and tools for unprecedented levels of coordination, cooperation, and competent governance.

But it would also pose a formidable range of challenges — challenges we need to solve to enjoy the fruits of the intelligence explosion. Such *grand challenges* are the main focus of our paper. One widely discussed challenge is the risk of AI takeover: that sufficiently advanced AI might be misaligned with and able to overpower humanity. But there are many other grand challenges, too, including:

- The risk of *human* takeover, by whoever controls superintelligence;
- Novel risks from destructive technologies, such as bioweapons, drone swarms, nanotechnology, and technologies we haven't yet concretely anticipated;
- The establishment of norms, laws, and institutions around critical ethical issues like the rights of digital beings and the allocation of newly-valuable offworld resources;
- The challenge of harnessing AI's ability to improve society's collective epistemology, rather than distort it.

Grand challenges such as these may meaningfully determine how well the future goes. They are forks in the road of human progress. In many cases, we shouldn't expect them to go well by default — not unless we prepare in advance.

Because of the accelerated pace of change from the intelligence explosion, the world will need to handle these challenges *fast*, on a timescale far shorter than what human institutions are designed for. By default, we simply won't have time for extended deliberation and slow trial and error.

Aligned superintelligence could solve some of these challenges for us. But not all of them. Some challenges will come before aligned superintelligence does, and some solutions, like agreements to share power post-AGI, or improving currently slow-moving institutions, are only feasible if we work on them before the intelligence explosion occurs.

In fact, there are ways we can prepare for these challenges, today. For example, we can:

1. **Prevent extreme and hard-to-reverse concentration of power**, by establishing institutions and policies now. For instance, we can ensure that data centers and essential components of the semiconductor supply chain are distributed across democratic countries and that access to frontier AI continues to be available to many parties, both within and across countries.
2. **Empower responsible actors:** Increase the chance that those actors who wield the most power over the development of superintelligence (such as politicians and AI company CEOs) are responsible, competent, and accountable.

---

7   We also do not imply that the field of "AI governance" has to date been exclusively concerned with misalignment risk (see [this research agenda](#) for illustration).

3. **Build AI tools to improve collective decision-making.** We could begin building, testing, and integrating AI tools which we want to be in wide use by the time the intelligence explosion is underway, across epistemics, deal-making, and decision-making advice.

4. **Remove obstacles to applying superintelligent AI to downstream challenges**, for example by unblocking bureaucracy which makes it unnecessarily difficult to use advanced AI tools in government.

5. **Get started early on institutional design for new areas of governance**, including on the rights of digital beings and the legal framework for property claims on offworld resources.

6. **Raise awareness and improve our understanding** of the intelligence explosion and each of the challenges that follow from it, so that it takes less time to get up to speed at the crucial moments when decisions do need to be made.

In the next section, we present a historical thought experiment to illustrate how dramatic and disruptive an intelligence explosion would be. In the sections following that, we:

- Argue that AI that can substitute for human researchers would drive a century in a decade of technological progress (Section 3)

- Give an overview of the wide range of grand challenges that would result (Section 4)

- Discuss when preparations for these challenges can and cannot be punted to a later time (Section 5)

- Present some of the things that we can do, today, to prepare for them (Section 6)

# 2. A century in a decade

## The last century, compressed

Consider all the new ideas, discoveries, and technologies we saw over the last century, from 1925 to 2025:

> **Scientific and conceptual**
>
> Advances in physics, including quantum theory and modern cosmology. Uncovering the structure of DNA and subsequent advances in genomics and biology. The theory of computation and computer science, Gödel's incompleteness theorems, cryptography, optimization and linear programming, and new statistical methods. Major advances in then-nascent social sciences like game theory, macroeconomics and econometrics, political science, experimental psychology, and forecasting. A much-improved understanding of environmental degradation.
>
> **Technological and engineering**
>
> Assault rifles, jet aircraft, radar and stealth technology, satellite reconnaissance, biological and chemical weapons, unmanned drones, and the atomic bomb. Penicillin, oral contraceptives, DNA fingerprinting, rational drug design, and genetic engineering. Transistor radios, digital computers, cell phones, the internet, social media, GPS, and deep learning. Fracking, offshore drilling, national electricity grids, nuclear power, and solar power. Synthetic fertilisers, high-yield crop strains, mechanisation with tractors and combines, and genetically-modified crops.
>
> **Political and philosophical**
>
> Uptake of secular and universal human rights, second- and third-wave feminism, decolonisation, animal rights, and environmentalism. The spread of fascism, state communism (such as Maoism and Leninism), neoliberalism, and 'realism' in international relations. New theories of political and social justice, new theories of normative ethics, cost-effectiveness approaches to appraising policy, and population ethics. Logical positivism, ordinary language philosophy, existentialism, critical theory, (post)structuralism, (post)modernism, and more.

Now, imagine if all of those developments[8] were instead compressed into the *decade* after 1925.

The first nonstop flight across the Pacific would take place in late 1925. The first footprints on the moon would follow less than four years later, in mid-1929. Around 200 days would have separated the discovery of nuclear fission (mid-1926) and the first test of an atomic bomb (early 1927); and the number of transistors on a computer chip would have multiplied one-million-fold in four years.[9]

These discoveries, ideas, and technologies led to huge social changes. Imagine if those changes, too, accelerated tenfold. The Second World War would erupt between industrial superpowers, and end with the atom bomb, all in the space of about 7 months. After the dissolution of European colonial empires, 30 newly independent states and written constitutions would form within a year. The United Nations, the IMF and World Bank, NATO, and the group that became the European Union, would form in less than 8 months.

Or even just consider decisions relating to nuclear weapons. On a 10x acceleration, the Manhattan Project launches in October 1926, and the first bomb is dropped over Hiroshima three months later. On average, more than one nuclear close call occurs per year. The Cuban Missile Crisis, beginning in late 1928, lasts just 31 hours. JFK decides how to respond to Khrushchev's ultimatum in 20 minutes. Arkhipov has less than an hour to persuade his captain, falsely convinced war had broken out, against launching a nuclear torpedo. And so on.

---

8   We'll sometimes refer to this as 'technological progress': using the term in a very broad sense, to refer to new technologies, feats of engineering, scientific breakthroughs, and other intellectual advances.

9   (Any 4-year period in the 1930s.)

Such a rapid pace would have changed what decisions were made. Reflecting on the Cuban missile crisis, Robert F. Kennedy Senior, who played a role in the negotiations, wrote: "If we had had to make a decision in twenty-four hours, I believe the course that we ultimately would have taken would have been quite different and filled with far more risks."

## Asymmetric acceleration

If literally every historical process accelerated uniformly, then the same events would just happen faster, and the trajectory of history wouldn't change. But technological change doesn't speed up everything equally: accelerating technological progress is like *slowing* the processes which don't accelerate along with it. In particular, it's hard to speed up humans themselves, and their interactions, thoughts, and learning. Similarly, many social and political institutions have rigid schedules,[10] which could drag behind as the rest of the world speeds up around them.

So an alternative thought experiment would be if the period 1925-2025 saw the same magnitude of technological change, but all human beings were awake for only [one hour and forty minutes per day]. Decision-making would clearly be impaired: with less time to understand new developments and their implications, but more pressure to act quickly, the odds of a major fumble would go up.

## The accelerated decade

Now imagine all the scientific, intellectual and technological developments that you would expect to see by the year 2125, if technological progress continued over the next century at roughly the same rate that it did over the last century. And then imagine all of those developments occurring over the course of just ten years.[11]

Like in the historical thought experiment, the challenges posed by this acceleration would be enormous. But, in the next section, we'll argue this is *not* just a thought experiment: it is in fact likely that advanced AI will drive a hundred years' worth of technological development in less than a decade.

The speed-up in technological development won't be uniform, because abundant cognitive labour from AI would speed up some areas of research more than others. Progress would speed up more in domains which can be advanced through a priori reasoning or simulations, like mathematics, computer science and computational biology, while progress would be less explosive in areas where expensive or slow experiments are necessary, like high-energy particle physics or drug development.

---

10  For example, the rollout of COVID vaccines in the US was reportedly delayed by the FDA's slow schedules, [despite efforts to speed them up]. Among other things, a [key meeting] aimed at granting the Pfizer-BioNTech vaccine emergency use authorization happened almost a *month* after the results of the Phase 3 trials were [announced] — possibly because, as the FDA [website] explains, the FDA is "required to publish announcements of advisory committee meetings at least 15 calendar days before a meeting date in the Federal Register."

11  A century in a decade may well be an understatement. If so, then it's more apt to consider the most consequential intellectual developments from the last millennium: evolution by natural selection, atheism, liberalism, democracy, abolitionism, feminism, universal secular human rights, Protestantism, limited-liability companies, capitalism, state socialism and communism, and breakthroughs as important as the printing press, the steam engine, calculus, the laws of motion, and the scientific method itself.

As in our historical thought experiment, the quality and speed of high-stakes decision-making would not always keep pace with the rate of change. Superintelligent AI advisors might help significantly (in ways we'll discuss in Section 5), but good decision-making will still be bottlenecked by the speed of human brains and pace of institutional schedules, and the usual method of institutional learning via trial and error on the timescale of human lifetimes will be too slow.

We are not arguing that this accelerated rate of technological progress would be *worse* than business-as-usual rates of progress. But we *are* arguing that such rapid progress would pose distinctive challenges, largely because human decision-making would struggle to keep up. And that suggests we should be preparing now.

# 3. The intelligence explosion and beyond

We'll now turn to the arguments for thinking that advances in AI will drive much faster rates of technological development.

The argument, in brief, is this. Currently, total global research effort grows slowly, increasing at less than 5% per year . But total AI cognitive labour is growing more than 500x faster than total human cognitive labour, and this seems likely to remain true up to and beyond the point where the cognitive capabilities of AI surpasses all humans. So, once total AI cognitive labour starts to rival total human cognitive labour, the growth rate of overall cognitive labour will increase massively. That will drive faster technological progress.

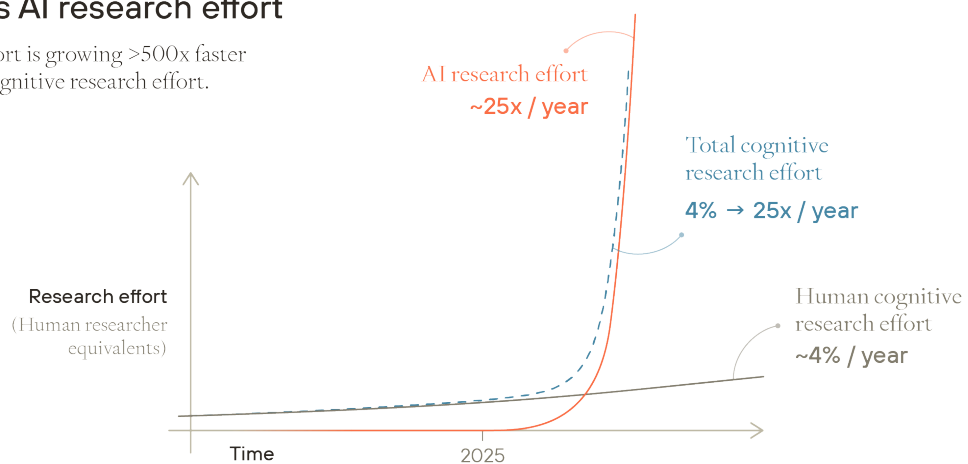

In this section, we lay out this argument in more detail. We look at the current and anticipated rates of AI progress, then discuss reasons for expecting AI to meaningfully accelerate research within a decade, and for thinking that the point at which AI reaches human parity might be soon.

We argue that, at this point, sustained rates of growth in research effort from AI are likely to be much higher than rates of growth in human research — enough to drive a century's worth of tech progress in less than ten years. Finally, we discuss why this would likely lead to explosive industrial expansion, too.

# Progress in AI capabilities

## Current trends

The best AI models are getting smarter, models at a given level of performance are getting more efficient, and we're accumulating more computing power to run them. The result is that total cognitive labour of AI systems is increasing dramatically every year.

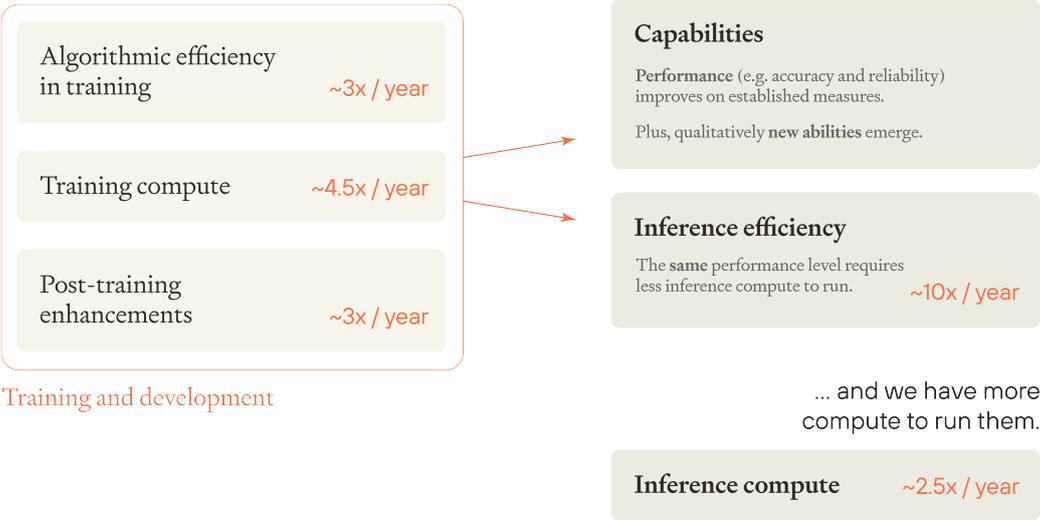

We'll go over the key drivers of AI progress in turn.

**Training compute:** The estimated amount of raw computation used in notable frontier training runs has been scaling up by around 4.5x per year since 2010.[12] Bigger training runs yield better models.

---

12   Sevilla, ‘Training Compute of Frontier AI Models Grows by 4-5x per Year’. Note that these numbers are drawn from a mixture of official and informal estimates, and uncertain informal estimates dominate among the largest recent training runs: the 90% confidence interval from Epoch.ai for the 2022–2024 trends spans 1.5x–11.8x annual growth. Such recent estimates may be partially influenced by other estimated trends, such as the rate of efficiency gains in training runs.

**Algorithmic efficiency in training:** New and better algorithms improve the gains from training, making the same amount of training compute more efficient. On current trends, the physical computation required to train a model at the same level of performance is falling by roughly 3x per year[13].

These two factors combine to give a measure of "effective training compute" — the equivalent increase in raw computation needed to reach the same model performance without any algorithmic innovation. Effective training compute from pretraining is increasing by over 10x per year.

**Post-training enhancements:** Adding to this, AI developers are finding ways to elicit new capabilities from the base (pretrained) models. They are building out "post-training enhancements" across areas like tool use, better prompting methods, synthetic data, creative ways of generating and selecting answers, and varieties of "scaffolding"[14]. Many of these advances make better use of more *inference* compute, essentially teaching trained base models to reason more effectively.[15] Anthropic informally estimates that post-training enhancements currently provide a *further* 3x efficiency improvement per year[16]. So, in terms of the capabilities of the best models, it's as if physical training compute is scaling more than 30x per year.[17]

These trends improve performance on familiar AI capabilities with established measures and benchmarks, often in predictable ways.[18] But qualitatively *new* capabilities emerge from bigger and better training runs, too, often in subtle and surprising ways. So it might be useful to consider the real-world difference between GPT-2 and GPT-4 and how that maps on to our trends. GPT-4 was trained using roughly $10^5$ (100,000) to $10^6$ (one million) times as much effective training compute as GPT-2[19], which was released around four years earlier. If we also factor in post-training enhancements, effective compute likely grew by closer to ten million-fold[20]. This led to

---

13  Ho et al., 'Algorithmic Progress in Language Models'. See also Erdil and Besiroglu, 'Algorithmic Progress in Computer Vision'. Note that this paper focuses on historical improvements, and does not argue at length about how they should be extrapolated.

14  Davidson et al., 'AI Capabilities Can Be Significantly Improved without Expensive Retraining'. Note that nearly all post-training enhancements studied here date from 2022-2023, suggesting we have less long-running historical data on post-training enhancements.

15  OpenAI, 'Learning to Reason with LLMs'.

16  Anthropic, 'Responsible Scaling Policy'. (p.16)

17  Note that this is based on expected *marginal* contributions from increases in scale (compute) and efficiency (algorithmic advances), since inputs to training are complementary; so we cannot straightforwardly extrapolate performance from gains in only one dimension over many orders of magnitude. For methodological discussion see Erdil and Besiroglu, 'Algorithmic Progress in Computer Vision'.

18  Owen, 'How Predictable Is Language Model Benchmark Performance?'.

19  The research institute Epoch AI estimates the training compute of GPT-2 (1.5B) and GPT-4 at 1.9E21 and 2.1E25 respectively; giving a scale-up of roughly 10,000x, with the 90% CI spanning roughly 5x in the latter case (Epoch AI, 'AI Benchmarking Dashboard'.) Using the central estimate of 3x improvements per year in pretraining algorithmic efficiency, and assuming 4 years' of improvement, gives a further ~80x multiplier on effective compute. But this is likely a slight overestimate, because (i) GPT-3 seemingly did not feature major algorithmic innovations on GPT-2 (Brown et al., 'Language Models Are Few-Shot Learners'.); and (ii) GPT-4 reportedly finished training by August 2022, locking in opportunities for gains in pretraining efficiency (OpenAI et al., 'GPT-4 Technical Report'.) So gains in pretraining algorithmic efficiency between GPT-2 and GPT-4 may have amounted to a 5–50x improvement, such that the increase in effective training compute between GPT-2 and GPT-4 likely falls somewhere between $10^5$ and $10^6$.

20  Most post-training enhancements were made between GPT-3 and GPT-4. Most notably, the GPT-3.5 range and onwards were optimised to be conversational assistants, using reinforcement learning from human feedback and other methods (Christiano et al., 'Deep Reinforcement Learning from Human Preferences'; Ouyang et al., 'Training Language Models to Follow Instructions with Human Feedback'). Since GPT-3 was qualitatively new capabilities: GPT-2 would produce grammatical but essentially meaningless text completions; GPT-4 was able to answer sophisticated questions about science, law, history, and coding.

**Example question**

The terms gas exchange, diaphragm, and inhale are most closely associated with which system in the human body?

**GPT-2** *(February 2021)*

Gas exchange is the process by which the body exchanges gases. Diaphragm is the term used to describe the opening of the chest cavity. Inhale is the term used to describe the process of inhaling air.

**GPT-4** *(March 2023)*

The terms "gas exchange", "diaphragm", and "inhale" are most closely associated with the respiratory system in the human body.

[Brief explanation of each term…]

Source: *AI Digest — 'How Fast Is AI Improving?'*

We can also run *more* model instances at the same level of performance, both thanks to more efficient models, and because AI companies are accumulating more computing hardware for inference. Taking both in turn:

**Inference efficiency:** GPT-3.5 was released in late 2022 at an initial cost of $20 per million tokens . Today, it is possible to run a faster and better model with a *better* score on MMLU (a wide-ranging benchmark for evaluating LLM performance) for around $0.04 per million tokens ; a 500x drop over less than 3 years[21] , or a yearly cost decline of around 10x per year . This figure has some extra theoretical support: doubling effective training compute very roughly corresponds

---

announced in May 2020, some 2.75 years before GPT-4, Anthropic's informal estimate of 3x per year from post-training enhancements would suggest roughly a further 20x gain in all-things-considered effective compute, with wide uncertainty.

21  OpenAI's text-davinci-003 175B model was released in late November 2022 , and achieved a score of 64.8 on MMLU (5-shot) (Chung et al., 'Scaling Instruction-Finetuned Language Models' ). Llama 3.1 8B was released in late July 2023 , and achieved a score of 73.0 on MMLU (Grattafiori et al., 'The Llama 3 Herd of Models' ). An archived image of openai.com/api/pricing shows inference costs of $0.02 per 1k tokens (input *plus* output tokens). At the time of writing Llama 3.1 8B can be run (using the cloud provider Nebius ) for $0.02 per million input tokens and $0.06 per million output tokens. As early as September 2024 , inference costs were $0.05 per million input tokens and $0.08 per million output tokens.

22  Jones (2021) finds a roughly linear trade-off in train-time compute with test-time compute in a simple board game: "for each additional 10× of train-time compute, about 15× of test-time compute can be eliminated". In the context of using more training to improve efficiency by training language models with smaller parameter counts, then we'd expect training efficiency to grow twice as quickly as runtime efficiency (in log compute). Comparing models using similar amounts of training compute to GPT-3, Hoffman et al. (2022) find that it is best to scale training data and parameter count roughly equally with a fixed training compute budget. So a language model can be made roughly $X$ times more efficient by reducing the parameter count by a factor of $X$ , but this increases training efficiency by a factor of $X^2$ (since training compute scales proportionally with data and parameter count, which are being reduced by the same factor). Thus a 2x increase in runtime to halving inference costs for an established capabilities level[22] , and we saw that effective training compute is increasing by around 10x per year.

**Inference compute scaling:** Increases in computing hardware work multiply these efficiency gains: the amount of physical compute available for inference is increasing at a rate of very roughly 2.5x per year.[23]

These two trends — runtime efficiency gains and increases in available inference compute — could then support growing the "AI population" by about 25x per year.[24]

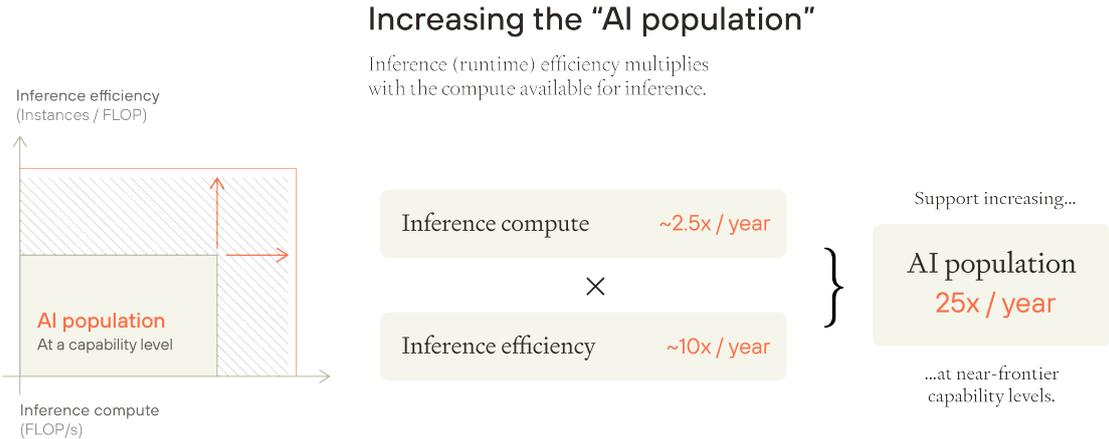

Increasing the "AI population"

Inference (runtime) efficiency multiplies with the compute available for inference.

All this adds up to rapid advances, not just in the most impressive capabilities of the best AI models, but in the *collective* capabilities of all the AI systems running in the world. We'll use *AI research effort* to talk about the quantity of research-relevant cognitive effort contributed by all active AI systems at a time. Informally, we can measure AI research effort in terms of the number of remote human researchers that would be needed to make an equivalent contribution.[25]

---

efficiency would require a 4x increase in effective training compute. See Jones, 'Scaling Scaling Laws with Board Games' , Hoffmann et al., 'Training Compute-Optimal Large Language Models' .

23  Total deployed accelerator FLOP/s (from NVIDIA chips) has grown by roughly 2.3x per year since 2019 (Frymire and Owen, 'The Stock of Computing Power' .) This does not count Google's TPU chips, older chips being repurposed for inference, or new chips specialized for inference in personal computers and phones. Note we are counting internal inference (such as for experiments) here, since some fraction of accelerator FLOP/s are only deployed internally. There are known tradeoffs between inference and training compute, such that the optimal allocation between inference and training compute is a balance; and empirical estimates suggest the balance is around even, unless methods to spend on training to save on inference (and vice-versa) effectively run out of steam (Erdil, 'Optimally Allocating Compute Between Inference and Training' ). If the frequency of large training runs does not change, and the training-inference balance does not change, this suggests that compute used for inference is also increasing by around 2.3x per year, with wide uncertainty. We can also note that revenue for major AI companies is growing by around 2–3x per year . We can assume that API costs include a fixed margin over hardware rental costs, and note that the price-efficiency of computation (in FLOP/s) per dollar cost new GPUs (amortised over two years) is increasing by around 1.35x per year thanks to Moore's law (Hobbhahn, 'Trends in Machine Learning Hardware' ). This again suggests around a 2×1.35 = 2.7x increase in inference per year. Recent advances in inference scaling , and the ability to repurpose a wider range of chips for inference, would suggest 2.5x is a slight underestimate.

24  In other words, if we devote all the growth in inference compute next year to running more efficient models, while holding their capabilities fixed, we could run 25x as many instances.

25  We can say total AI and human cognitive research effort are at parity if the contribution to overall cognitive research outputs from AI is equivalent to instead effectively doubling the number of human researchers. So we can quantify AI research effort in terms of the equivalent number of "effective remote-working human

Total human cognitive research effort is a function of the sheer number of researchers, and the average productivity of each researcher. Similarly, we can think of AI research effort as a function of the capabilities of each model, and the amount of "AI researchers".

Suppose that current trends continue to the point where AI research effort is roughly at parity with human research labour (below, we discuss whether this is likely). For the sake of argument we might picture the same effective number of human-level "AI researchers" as actual human researchers . How radically would this affect the growth rate in *overall* cognitive labour going towards technological progress?

We can estimate a baseline by assuming that progress in training is entirely converted into improved inference efficiency, and more inference compute is entirely used to run more AIs.[26] We have seen that inference efficiency is improving roughly in line with effective training compute at around 10x per year, and inference compute is increasing by 2.5x per year or more. So *if* current trends continue to the point of human-AI parity in terms of research effort, then we can conclude AI research effort would continue to grow by at least 25x per year.[27]

But this likely underestimates the growth rate in AI research effort. Above we assumed that doubling the effective training compute available for AI-driven research would exclusively be used to increase the *population* of AI researchers, but it could instead allow us to run fewer but *smarter* models. If more effective training compute results in models twice as efficient for the same capability level of the last generation, and models with new capabilities which use the same inference compute, the same number of smarter AIs are often going to be far more collectively capable than twice the number of equally smart AIs. Similarly, doubling inference compute can be used to double the serial reasoning time *per* AI researcher,[28] instead of just running more researchers. And of course any mix of both factors is possible. We should expect the gains from training and inference to be applied where they are most productive, and we can also just observe that practical applications of AI do in fact tend to use models close to the public frontier in capabilities.

We don't make estimates for how quickly AI research effort would grow when we account for capabilities improvements beyond human level, but it could be *much* faster. To see this, consider that human researchers vary only modestly in brain size and years of education, yet in some fields

---

researchers" needed to make the same contribution absent frontier AI. We can say that AI research effort has further doubled if its contribution measured in terms of effective humans researchers again doubles from this point of human-AI parity, and so on.

26. That is, we assume the fraction of inference used for research stays roughly constant, and then assume that expansions in inference compute are used to run more instances, rather than to scale the inference compute used per instance.

27. This assumes that AI research effort (as defined in terms of equivalent cognitive contributions from human researchers) increases in line with the population of AI researchers. We think this is a reasonable assumption but there is room for disagreement here.

28. Recent advances have shown how pretrained models can use more serial inference compute to 'reason' through a chain of thought (OpenAI, 'Learning to Reason with LLMs' ). But there are simpler methods, too, like many-shot in-context learning (Agarwal et al., 'Many-Shot In-Context Learning' ), MCTS methods, and where verification is relatively cheap, simply making many attempts and picking the best result. (Villalobos and Atkinson (2023) ) conclude that, at the optimal inference-training balance, "we expect that each technique makes it possible to save around 1 order of magnitude (OOM) of compute in either training or inference, in exchange for increasing the other factor by somewhat more than 1 OOM." (Villalobos, 'Trading Off Compute in Training and Inference' ).

exceptional researchers can be hundreds of times more productive than their peers.[29]

We have also not yet accounted for the purported 3x per year increase in effective training compute from post-training enhancements, which would increase the growth rate from 25x to 75x per year. Combined with the extra gains when we account for the option of making AI researchers *smarter* instead of just more efficient, the growth rate in total AI research effort could far surpass that number.[30]

## How far could AI keep improving?

For how long can these trends in scaling and efficiency continue?

- **Training compute:** The biggest training runs can continue to scale for roughly another 10,000x increase before running into power and other limits (likely well within a decade from now). Even if that barrier were solved, training runs could hit limits from chip production, data scarcity, and hardware latency.[31]

- **Algorithmic efficiency in training:** If we assume the same ratio of compute scale-up to efficiency advances, we'll see a further 1,000x increase in algorithmic efficiency in training,[32] giving a further ten million-fold increase in effective training compute over a decade, compared to where we are today — or about a 5x yearly increase in effective compute over the decade — slower than the ⩾10x yearly rates of progress to date.

- **Inference compute:** We can reasonably assume that inference compute would *also* scale by 10,000x over this period, at a continued 2.5x yearly increase.

So if we conservatively assume AI progress continues as far as scaling carries it, and no further, we can use the limits on scaling to estimate how much collective AI research effort could grow over the coming decade. The product of training compute, algorithmic efficiency, and inference compute — which combine to give AI research effort — will then have increased *one hundred billion-fold* ( $10^{11}$ ), averaging just over 10x per year.

---

29     The most obvious measure here is the distribution of citations across various fields, though we can also note the high salaries of exceptional researchers in fields like AI. For more, see ['[EA Forum] How Much Does Performance Differ between People?'](#) . We can also just look at anecdotes of how improvements to capabilities can be equivalent to a vast multiplier on the number of people. For example: in 1999, the strongest chess player at the time — Garry Kasparov — played a game of chess against the "world", where four world-class chess players suggested move options and tens of thousands of participants analysed and voted on their response. Kasparov won.

30     3x yearly increases in effective compute from post-training enhancements brings growth in effective training compute to 30x yearly, bringing the growth rate in AI research effort to 75x. But there are also often far greater gains from upgrading model capabilities, rather than running the same capabilities more compute-efficiently. This would take the growth rate in AI research effort beyond 75x.

31     Pilz, Mahmood, and Heim, ['AI's Power Requirements Under Exponential Growth'](#) ., Sevilla, ['Can AI Scaling Continue Through 2030?'](#)

32     Physical training compute is currently growing by 4.5x yearly, and algorithmic efficiency in training is improving by around 3x yearly. So for every 10x increase in physical compute, algorithmic efficiency improves by approximately 5.4x ( $= 3^{log(10)/log(4.5)}$ ). A 10,000x scale-up in physical compute is four 10x increases, giving a factor improvement in algorithmic training efficiency of 835x ≈ 1000x.

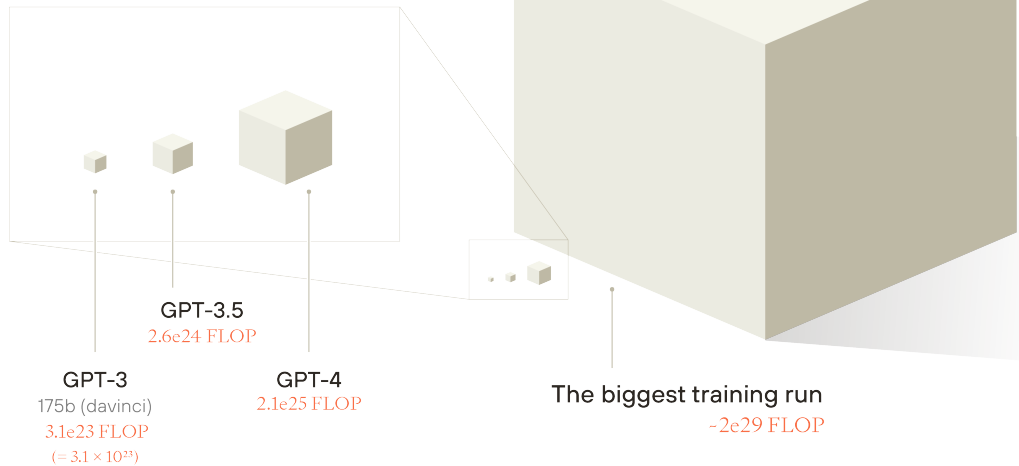

### Scaling the biggest training run

Scaling could support a 10,000x increase in the compute used for the biggest training run over the coming decade, before running into power and other constraints.

GPT-3
175b (davinci)
3.1e23 FLOP
(= 3.1 × 10²³)

GPT-3.5
2.6e24 FLOP

GPT-4
2.1e25 FLOP

The biggest training run
~2e29 FLOP

Source: Epoch AI, 'Data on Notable AI Models', 'Can AI Scaling Continue Through 2030?'

But while physical scale-ups will likely run into limits within a decade, algorithmic advances could dramatically *accelerate*. This is because AI models could themselves begin meaningfully accelerating improvements in AI algorithms, data, post-training enhancements, and other software techniques. That would result in better AI, which can then drive algorithmic improvements, and so on: a *software feedback loop*,[33] where AI capabilities keep improving without any need to scale up physical inputs.

A self-sustaining software feedback loop isn't guaranteed. Most new technologies are (at best) a one-off multiplier on researcher productivity, so they can't cause self-sustaining progress.[34] In principle, things could be similar in AI. Even if "digital ML researchers" could substitute for a typical human AI researcher, progress could level off: the first generation of digital ML researchers could find ways to double the performance of the next generation of AI researchers, but the v2 digital ML researchers might only find ways to improve performance by 1.5x, and so on: the result would be a flash of improvement that decelerates as soon as it begins. For progress to be rapid and sustained, a doubling of inputs must yield at least a doubling of outputs, over many doublings.

However, empirical estimates of efficiency gains in various software domains suggest that doubling cognitive inputs (research effort) generally yields more than a doubling in software performance or efficiency.[35] Based on this, reasonable estimates suggest a roughly even chance that an AI-driven software feedback loop *would* drive accelerating progress in AI performance and efficiency.[36]

---

33   Davidson, Hadshar, and MacAskill, 'Three Types of Intelligence Explosion'

34   Consider the photocopier. The first successful commercial plain paper copier was the Xerox 914, released in 1959. Now researchers could copy text, saving trips to the library, and so the Xerox 914 lastingly (if insignificantly) sped up aggregate research progress. Moreover, the speed-up surely boosted R&D into the next generation of photocopiers, since Xerox employees themselves could now work faster, and so on. So the Xerox 914 technically triggered a positive feedback loop in photocopier R&D, but it fizzled out almost instantly, rather than causing explosive progress in photocopier technology.

35   Erdil, Besiroglu, and Ho, 'Estimating Idea Production'

36   Davidson, Hadshar, and MacAskill, 'Once AI Research is Automated, Will AI Progress Accelerate?'

Feedback loops can't accelerate forever, so eventually the software feedback loop has to plateau. But the upper bound on algorithmic efficiency seems *very* far from current levels. LLMs seem to learn very roughly 100,000 times less efficiently than humans, in terms of the computation our brains use by adulthood.[37] And the upper bound seems likely to be far *beyond* the efficiency of human brains.[38]

So, if we do get a software feedback loop, AI capabilities could cover *another* factor of one million-fold in effective training compute within the same decade, giving roughly a trillion-fold ($10^{12}$) increase in effective training compute, or a ten quadrillion-fold ($10^{16}$) increase in the product of inference compute and effective training compute (roughly 40x per year).

Either way, the product of inference compute and effective training compute is growing at least 600 times faster[39] than all human cognitive labour devoted to technological progress, with a vast amount of headroom for continued improvements.

## Estimates of AI progress over the coming decade

In the table below, we summarise our guesses for how fast components of AI research effort will continue to grow, on average, over the next decade. Note that given growth rates are averages — most growth in a given factor could be very concentrated in a short period of time.

|  | Current rates | Moderate Scenario (No feedback loop) | Rapid Scenario (Software feedback loop) |
| --- | --- | --- | --- |
| Training compute | 4.5x/yr | 2.5x (10,000x total) | 2.5x |

---

37  Cotra (2021) estimates that human lifetime learning uses the equivalent of roughly $10^{24}$ FLOP of computation. Assuming that training compute continues to scale at historical rates, GPT-6 will be trained with around $10^{29}$ FLOP. If GPT-6 were as capable as a human being, then it would be 100,000 times less training-efficient than human learning. So a software explosion starting with GPT-6 could span a 100,000x gain in efficiency. Of course, the software feedback loop could hit diminishing returns before reaching the absolute ceiling on algorithmic efficiency. But if we are (for example) uniformly uncertain about where the plateau is located over all possible doublings, then the absolute ceiling is nonetheless a useful guide to the expected number of doublings a software feedback loop will drive.

38  Humans are efficient learners, but biological brains are limited in the algorithms they can implement, and we can't directly interface with computer tools. For example, our brains cannot implement weight sharing (Ott et al., 'Learning in the Machine'). Human brains also seem to be drastically undertrained. If the optimal tradeoffs between compute and training data in AI systems carry over to brains, then compute-optimal learning with our brains would involve learning from roughly 10,000x more data than we see in a lifetime. See: Davidson, Hadshar, and MacAskill, 'How Far Can AI Progress before Hitting Effective Physical Limits?'; Hoffmann et al., 'Training Compute-Optimal Large Language Models'. For most tasks, then, optimal learning and inference algorithms running on equivalently powerful hardware would probably be *far* beyond 10x as efficient as a biological brain.

39  The supply of effective human researchers is growing on aggregate by at most 4% per year (Bloom et al., 'Are Ideas Getting Harder to Find?'), although this growth rate is ultimately limited by human population growth, currently around 1% and declining. We argued that marginal growth in cognitive labour (for research or otherwise) is roughly proportional to growth in effective training compute (currently 10x/year) and inference compute (currently 2.5x/year), which combine to give 25x/year. 25x growth is 600 times faster than 4% growth.

| | | | |
|---|---|---|---|
| Algorithmic efficiency | 3x/yr | 2x (1,000x total) | 8x ($10^9$ x total) |
| Inference compute scale-up | 2.5x/yr | 2.5x (10,000x total) | 2.5x (10,000x total) |
| AI research effort | ⩾25x/yr | ⩾12x (⩾ $10^{11}$ x total) | ⩾50x (⩾ $10^{17}$ x total) |
| Human researchers | 4% | 4% | 4% |
| AI vs human research effort | ⩾ 600x faster | ⩾ 300x faster | ⩾ 1000x faster |

Putting this all together, we can conclude that *even if current rates of AI progress slow by a factor of 100x compared to current trends*, total cognitive research labour (the combined efforts from humans and AI) will still grow far more rapidly than before. Once collective AI capabilities match that of humans, this growth would continue to take AI capabilities far beyond the collective thinking power of all humanity.

Next, we turn to the question of *whether and when* collective AI capabilities will reach parity with human research effort.

## AI-human cognitive parity

It seems likely to us that AI research effort will reach parity with human research labour within the next two decades, meaning AI systems can collectively perform almost all the research-relevant cognitive work that humans can. Because scaling is driving such large gains, AI could even approach human parity well within the coming decade, before scaling begins hitting power and other practical constraints. And even after scaling slows down significantly, algorithmic progress could keep driving progress forward. In any case, AI-human parity could arrive in a matter of years.

To see this, we should look directly at how AI capabilities are improving. On GPQA — a benchmark of Ph.D-level science questions — GPT-4 performed marginally better than random guessing. 18 months later, the best reasoning models outperform PhD-level experts.[40]

---

40  Those models include DeepSeek's r1 model, which was reportedly trained using roughly 4x *less* compute than GPT-4. (Epoch AI, 'AI Benchmarking Dashboard').

### AI performance on a set of Ph.D.-level science questions
*GPQA Diamond accuracy*

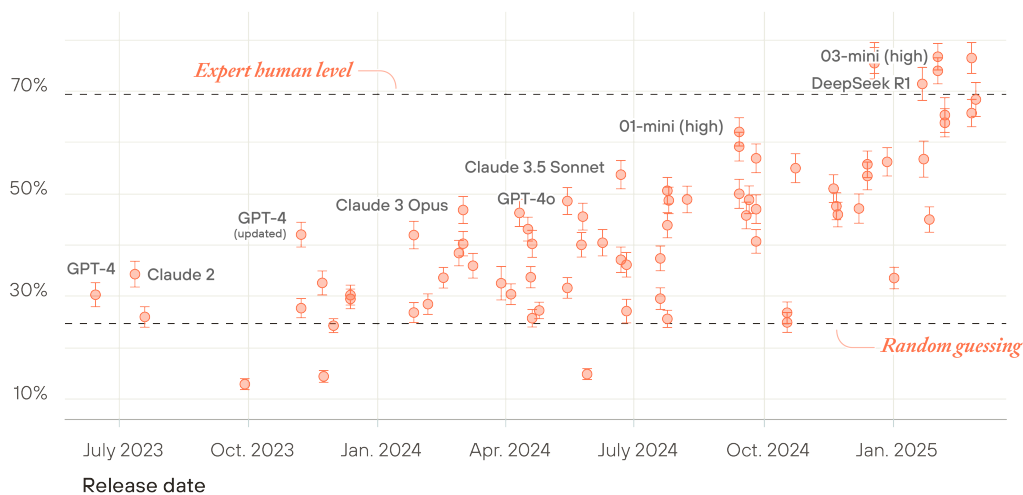

Source: *Epoch AI — AI Benchmarking Hub (2025)*

As for skills more relevant for automating ML research: the best AI systems can already solidly beat all but the very best programmers on competitive coding questions,[41] solve more than 70% of real-world software engineering issues in open source repositories,[42] and match domain experts' scores on time-capped machine learning optimisation problems, up to a time limit of around four hours (on the RE-Bench benchmark).[43] The best systems today fall short at working out complex problems over longer time horizons, which require some mix of creativity, trial-and-error, and autonomy. But there are signs of rapid improvement: the maximum duration of ML-related tasks that frontier models can generally complete has been doubling roughly every seven months. Naively extrapolating this trend suggests that, within three to six years, AI models will become capable of automating many cognitive tasks which take human experts up to a month.[44]

So at least on some benchmarks that test performance on short science and coding tasks, AI systems are catching up to that of humans with relevant expertise.

Of course, benchmarks are not the real world. Take ML research as an example: even tough benchmarks like RE-Bench present self-contained, well-defined tasks with quick feedback loops.

---

41  OpenAI et al., 'Competitive Programming with Large Reasoning Models'. On the popular competitive coding website Codeforces, only around 200 active users remain with a higher rating than the best AI coders. One user wrote: "I knew it would surpass me eventually, but I didn't expect it to happen so quickly—just about a year. Now, I feel slightly panicked".

42  This was mentioned in an official OpenAI livestream announcing o3 and o3-mini, available to watch here. "On software style benchmarks, we have SWE-Bench Verified, which is a benchmark consisting of real-world software tasks. We're seeing that o3 performs at about 71.7% accuracy, which is over 20% better than our o1 models."

43  Wijk et al., 'RE-Bench'.

44  METR, 'Quantifying the Exponential Growth in AI Ability to Complete Longer Tasks' (forthcoming). See also Pimpale et al., 'Forecasting Frontier Language Model Agent Capabilities'.

While in the real world, ML researchers need to clarify and choose between research directions, and coordinate with other teams, while often waiting days for experiments to resolve.[45] It will take a lot more work to build and integrate AI systems which can autonomously drive forward actual research and engineering, though there are huge incentives to cross the gaps between expert-level benchmark performance and expert-level usefulness in the real world.

Around a millionfold increase in effective training compute separates GPT-2 — which would give barely coherent answers to basic questions — and GPT-4, which was a useful assistant with more general knowledge than any living person.[46] In 18 months, models improved enough to match PhD-level experts on certain reasoning and programming tasks related to their fields. We should expect at least the same leap in algorithmic progress and training compute that separated GPT-2 and GPT-4 before we run up against bottlenecks to scaling.[47] If we get a software feedback loop, we could get roughly the same leap *again* , getting us to the equivalent of "GPT-8" and then some.[48]

Such intense progress seems fairly likely to us to be sufficient to produce models which surpass the research ability of even the smartest human researchers, in basically every important cognitive domain. But even if the expected scale-up in training compute over the next five years doesn't get us there, pre-training algorithmic efficiency, post-training enhancements, inference efficiency and scale up in inference compute would keep increasing AI research effort beyond the point where physical scaling slows down, or training data dries up.

So, though we cannot know for certain when it will happen, AI parity with the best human performance at research within the coming years seems more than a live possibility to us — it seems fairly likely. And when that happens, collective AI research capability will *continue* to increase across many orders of magnitude.

So we might start with an effective population of, say, one hundred thousand human expert-level AI researchers. Perhaps most will be working on improving the next generation of AI, or other areas of science and technology — wherever returns are highest. Then effective compute advances and increases in inference compute could at least double the amount of cognitive effort, and double it again, rapidly outsmarting all roughly ten million[49] working human researchers, then all 8 billion living humans, before then doubling in collective capabilities many times over again.

Of course, specialised AI applications are already driving progress in specific domains far more effectively than adding additional human researchers — for example, Google DeepMind's AlphaFold has been used to predict more than 200 million new protein structures,[50] in some ways achieving what would have taken experimental methods millions of researcher-years.[51] We won't

---

45   Owen, ['Interviewing AI Researchers on Automation of AI R&D'](#) .

46   AI Digest, ['How Fast Is AI Improving?'](#)

47   At current rates of progress, such a leap would take just over four years. Remember the estimate of 30x per year increase in effective training compute, which includes post-training enhancements. As discussed, the gap in effective training compute between GPT-2 and GPT-4 was at most a factor of one million in effective training compute. At this rate, a further million-fold leap in effective compute takes just over four years.

48   GPT-5 is unreleased at the time of writing, but the best released models are qualitatively around half a GPT-gap above GPT-4 on release.

49   Ayan, Haak, and Ginther, ['How Many People in the World Do Research and Development?'](#)

50   Jumper et al., ['Highly Accurate Protein Structure Prediction with AlphaFold'](#) ., Yang et al., ['AlphaFold2 and Its Applications in the Fields of Biology and Medicine'](#) .

51   Note that AlphaFold does not fully solve the problem of predicting the biological properties of a given protein and explaining why it folds the way it does. It primarily outputs final protein structure predictions, which can be limited to only using general-purpose AI researchers with the same skill profile as human researchers.

## How much more AI research effort after AI-human parity?

We can ask, then, how much *more* AI research effort will grow in the decade after it approaches parity with human cognitive research efforts? This depends on two guesses: how close AI capabilities already are to reaching parity with humans, and how much headroom you think there is for sustained and rapid improvements in overall AI capabilities.

In our **conservative** scenario, we'll imagine that AI capabilities reach human parity at a time when further scale-ups of training runs are no longer possible due to power constraints; and that there is no software feedback loop. What's more, we'll suppose that inference compute growth and efficiency advances slow by about 30% from their present rates. In this scenario, AI research effort would continue scaling, on average, at 5x per year for the decade following human parity.

In our **aggressive** scenario, we'll imagine that collective AI capabilities reach human parity soon, with plenty of room still to go in scaling up training runs and inference compute, and that AI automating AI R&D *does* result in a software feedback loop.

|  | Conservative | Aggressive |
| --- | --- | --- |
| Training compute | No growth | 2x (1,000x total) |
| Algorithmic efficiency | 2.5x (10,000x total) | 5x ($10^7$ total) |
| Inference compute scale-up | 1,000x (2x/yr) | 2.5x (10,000x total) |
| AI research effort | ~ $10^7$ x (5x/yr) | 25x (~ $10^{14}$ x total) |
| Human researchers | 1.04x/yr | 1.04x/yr |
| AI vs human research effort | 100x faster | 600x faster |

# The technology explosion

The next question is what happens to technological progress. The speed of technological progress has historically been driven and limited by the supply of skilled human researchers. Institutional

---

also be determined by letting proteins fold experimentally and analyzing the results. Experimental methods, unlike AlphaFold, also enable a finer analysis of side ligands and other properties essential to drug function and practical applications. With all these caveats in mind, at the current rate, we estimate it would have taken 80 years and 300 billion research-hours for experimental methods to output what AlphaFold did in one year. Moreover, as researchers figure out how to use and improve it, AlphaFold's biggest contributions are likely still in the future.

and cultural factors matter too, but few would deny that (all else equal) more researchers produce more research.[52] By increasing the overall amount of *cognitive* research effort, AI research effort could effectively grow the supply of skilled human researchers. If the effect is large enough, the result could be a *technology explosion*: a large and sustained increase in society's rate of technological progress, perhaps enough for a century in a decade of technological progress.

Any modeling of the effects of increases in total research effort needs to take into account two types of diminishing returns. First, the "fishing out" effect: across virtually every field studied, the more cumulative technological progress is made, the more research effort is needed to sustain the same rate of progress.[53] Second, the "stepping on toes" effect: there are often diminishing returns to adding more research effort in parallel.[54]

We model this via a simple idea production function from semi-endogenous growth models[55], and incorporate the two types of diminishing returns. In order to drive a century in a decade, total research effort would need to increase by a factor of 600 or more over that decade, that is, with a growth rate of around 100% (one doubling) per year.[56]

On our conservative scenario of AI progress over the post-human-parity decade, total research effort would increase by a factor of $10^7$, or about 5x per year. This would be enough to drive over

---

[52] For supporting models, see Jones, 'The Past and Future of Economic Growth'; Kremer, 'Population Growth and Technological Change'. For suggestive empirical findings, see Moser, Voena, and Waldinger, 'German Jewish Émigrés and US Invention'; Borjas and Doran, 'Productivity of American Mathematicians'.

[53] The most obvious explanation is that we find the easy-to-find ideas first, so the ideas left over are harder to find. Bloom et al. estimate that, between 1930 and 2020, a measure of the "effective number of researchers" worldwide increased by a factor of roughly 20, while the rate of technological progress remained roughly constant. (Bloom et al., 'Are Ideas Getting Harder to Find?')

[54] The most obvious reason is that when the number of researchers at a time increases, researchers are more likely to accidentally duplicate one another's work. It's hard to estimate the size of this effect, but a conservative estimate would suggest that increasing research effort by 100x over one year is equivalent to instead speeding up research by a factor of (only) around 30x over the same year. So in order to get another 100 years of technological progress in a decade, average research effort over that decade would need to increase far more than 10x current levels. In the following idea production function: $\frac{\dot{A}_t}{A_t} = \alpha S_t^\lambda A_t^{-\beta}$, this corresponds to λ = 0.75. Bloom et al. use this value, and Jones and Williams provide some evidence that λ > 0.5. (Jones and Williams, 'Too Much of a Good Thing?') Ekerdt and Wu (2024) argue that around half of what appears to be a "stepping on toes" effect is in fact due to a decline in the average ability of researchers, since the highest-ability potential researchers self-select into research first, and the fraction of people doing research has grown. This suggests λ would be closer to 1 for AI researchers (since there would be no such reason for additional AI researchers to decline in ability), and thus technological progress from AI would be faster than our estimates. ('resEW.Pdf'.)

[55] Jones, 'R&D-Based Models of Economic Growth'; Kortum, 'Research, Patenting, and Technological Change'; Segerstrom, 'Endogenous Growth without Scale Effects'.

[56] We can estimate this analytically. Again we can assume a default steady-state trajectory, which as we saw is given by: $\frac{\dot{A}_t}{A_t} = \alpha S_t^\lambda A_t^{-\beta} \Rightarrow g_a = \frac{\lambda g_S}{\beta} \Rightarrow g_a \propto g_S$ Assume, on the default path, that the number of researchers grows 4% yearly (driving 1.25% TFP growth where λ = ¾ and β = 2.4). The steady-state growth rate in researchers for 10x faster tech progress is then just 10x the default steady-state growth rate, that is 40% per year. But there would be some delay for the technology growth rate to converge to λ × gS / β. The initial value of St can be raised by a factor of $10^{1/\lambda}$, (≈ 21.5x) to instantaneously set the rate of technological progress to ten times the default level, and then 40% growth would maintain a steady-state of 10x sped-up technological progress for 10 years. The resulting increase over the original supply of human research is $10^{1/\lambda} \times (1 + (\lambda \times g_S/\beta))^{10} \approx 21.5 \times 140\%^{10} \approx 620 \times$. The average growth rate in the number of researchers is then around 90%, although the true growth rate required will be slightly higher, since we are frontloading research by assuming an instantaneous jump. Using numerical methods, we can confirm that the exact growth rate required is indeed around 100%.

*three hundred years'* worth of technological development on current rates of progress.[57] Even if per-decade growth in collective AI capabilities is less than half as slow as our conservative estimate suggests, we would still get a century in a decade of technological progress.

So far, we are assuming that *cognitive* effort from AI researchers can straightforwardly substitute for human research effort, and so the pace of technological progress is just a function of cognitive effort — of a priori, disembodied thinking. But research also involves physical experiments and tinkering. So unless physical labour available for research grows in line with cognitive labour, it will increasingly drag on technological progress. Similarly, technological progress often relies on waiting for experiments which have to be performed in serial[58] or can't easily be sped up, like human drug trials. And then there is physical capital, like lab equipment and pilot plants. Since capital complements labour in research, piling up more AI cognitive labour would become less and less useful.

These are good reasons to expect technological progress to speed up less than the naive model given above. But physical labour, experiments, and capital limitations are unlikely to so severely limit explosive technological progress that we would not see a century's worth of development in less than a decade. This is for several reasons.

First, overall progress at a time is given by average rates of progress; it's not constrained by the slowest-developing areas of technology.[59] Even if there were no progress at all in medicine or high-energy particle physics, we could still have a century's worth of development in a decade if there is enough progress in mathematics, computer science, theoretical economics, and simulation-heavy areas like computational biology. And even areas of research that currently rely heavily on capital and physical experiments would often still be able to make dramatic progress, given an enormous influx of cognitive labour. Those areas could still make major theoretical progress. AI researchers could look back on and re-use existing data. And high-quality simulations could often be used in place of physical experiments in some fields, such as synthetic biology, biochemistry, pharmacokinetics,[60] materials science,[61] robotics,[62] and even the social sciences.[63]

---

57  Consider this idea production function: $\frac{\dot{A}_t}{A_t} = \alpha S_t^\lambda A_t^{-\beta}$ Where $A_t$ is the technology level, St is is the number of researchers, $\alpha$ is a constant productivity parameter, the elasticity λ determines the rate at which research productivity declines with the number of researchers, and the elasticity β determines the rate at which research productivity declines with the technology level. Bloom et al. estimate β = 3.1 for the aggregate economy between 1930–2015, assuming no "stepping on toes" effects (i.e. assuming λ = 1). The steady-state growth rate (where growth in the technology level is constant) is then given by: $\frac{\dot{A}_t}{A_t} = g_a = \frac{\lambda g_S}{\beta}$ Estimates of λ for the aggregate economy are difficult, because they are hard to disentangle from diminishing returns from technological progress itself (i.e. from β). We follow the assumption of λ = ¾ from Bloom et al. Introducing λ = ¾ causes a downward revision to historical estimates of β, since λ = ¾ suggests that some fraction of diminishing researcher productivity is due to increases in the number of researchers, and hence a correspondingly smaller fraction is due to technological progress itself. So for λ = ¾, Bloom et al. revise their estimate of β from 3.1 to 2.4, for the period 1930–2015. We then use λ = ¾ and β = 2.4 to estimate rates of technological progress from more rapid growth in the number of effective researchers from AI. We define a "century in a decade" as a century in which technology improves at the equilibrium rate where the number of researchers continues to grow at average rates from the past century. Bloom et al. estimate the average growth rate of research effort from 1930–2015 as 4.3% per year, and we use 4% for simplicity. We assume technological progress in the aggregate economy is growing at a steady state where the growth in researchers is 4% (that is, 1.25% annual growth in TFP). Using our conservative estimate of 5x growth in effective researchers in the decade after human-AI research effort parity, and assuming no effect on technology previous to human-AI parity, we then find (numerically) the new TFP level after a decade of growth in AI research effort to be around 50x the level at the beginning of the decade, equivalent to more than 300 years of technological progress on the default (non-AI) growth path.

58  One way to model this is to imagine that a fast-growing AI researcher population could parallelize the more parallelizable tasks first, so that the parallelizability of the marginal research task falls with the researcher population. Eventually there is almost nothing useful to be done with more researchers; progress requires waiting on experiments which must be performed in series. In particular, we can replace our idea production function with $g_A = \theta \ln(S)$ — eliminating all "fishing out" effects, but where ideas accumulate logarithmically with $S$, instead of with a power function. Define the units of $S$ so that the current elasticity of $g_A$ with respect to $S$ is ¾. Then we find that multiplying $S$ by $10^7$ yields $g_A = 0.196$: somewhat higher than ten times the 1.5% growth rate in TFP that would count as a 10x increase in the growth rate, but close enough that the case no longer seems clear, especially once we add "fishing out" back in. This is somewhat closer to the model from Young (1998). Thanks to Phil Trammell for this point (errors in rephrasing it are ours).

Second, given an abundance of cognitive labour, we would see extremely optimised uses of physical capital, and extremely optimised experiments. If capital and physical experiments are the bottleneck, then it would be worth spending the equivalent of millions of years' worth of human research labour in order to design and run the most useful experiments possible, to collect huge amounts of data, and to analyse the data in intense detail. Relatedly, there would be very strong incentives to invest more in physical capital and to run more experiments. This could come about by first reallocating human labour (perhaps managed by AI scientists) towards running more experiments, often in parallel, and then later using robotics to automate the experiments.

Finally, as well as considering the limitations of a huge AI research effort force, we should consider the special advantages that AI systems would have, too. Evolution did not optimise human brains to be good at science and engineering; we are able to do science at all only by repurposing a general reasoning ability that evolved to help us survive in the ancestral environment.[64] In contrast, AI researchers could be intensely optimised for producing research contributions. Because they can run much faster than human brains, and run in parallel, AIs can undergo training that simply isn't possible within a single lifetime — spending the equivalent of a million years learning an area of mathematics that was discovered only five years ago. They could be experts in many different fields of science, and therefore much more able to bring insights from one field to bear on another. They could focus on a single problem for the equivalent of thousands of consecutive human lifetimes. And while computers already outshine humans at programmable tasks, there is no principled reason AI could not overtake the best humans in domains associated with sparks of insight, intuition, and creativity — like how Ramanujan or Einstein outshone graduate students with similar calculating ability.[65]

---

59   That said, you can't straightforwardly derive TFP progress across time from the average rates of progress at successive times, because (for instance) productivity advances can cause a sector to shrink if the price elasticity of demand is low.

60   'Revolutionizing Pharmacokinetics: The Dawn of AI-Powered Analysis'.

61   Zeni et al., 'A Generative Model for Inorganic Materials Design'.

62   Muratore et al., 'Robot Learning From Randomized Simulations'; Torne et al., 'Reconciling Reality through Simulation'.

63   Horton, 'Large Language Models as Simulated Economic Agents'.

64   Consider Bostrom (2014): "Far from being the smartest possible biological species, we are probably better thought of as the stupidest possible biological species capable of starting a technological civilization — a niche we filled because we got there first, not because we are in any sense optimally adapted to it.". Similarly: "One can speculate that the tardiness and wobbliness of humanity's progress on many of the "eternal problems" of philosophy are due to the unsuitability of the human cortex for philosophical work. On this view, our most celebrated philosophers are like dogs walking on their hind legs - just barely attaining the threshold level of performance required for engaging in the activity at all." (ibid.)

65   Consider that the general theory of relativity is one of the most impressive conceptual advances ever, but Einstein didn't develop it just by piling up empirical data. Instead, he combined deep understanding of the So, even taking into account headwinds arising from physical experiments and physical capital, we think a century of technological progress in a decade — or far more — is more likely than not.

Our naive model suggested that a roughly 1,000x increase in AI research effort would cause more than a century's worth of technological progress in a decade. Being conservative, let's assume that complications around physical experiments and capital mean we really require a further 10x increase in cognitive research effort over the course of 10 years — a 10,000x increase.[66] But current trends suggest that, over ten years, once AI reaches human parity we will get somewhere between a ten billion-fold increase in AI research capability (if compute scaling halts and even if algorithmic efficiency improvements somewhat slow down) and a hundred trillion-fold increase (if we get an aggressive software feedback loop). Even on the conservative projection, and even factoring the headwinds into account, there would be orders of magnitude more growth in collective AI cognitive effort than is needed to drive a century's worth of technological development in a decade.

Therefore, on a default path where we keep scaling AI without collectively agreeing to slow down, a century's worth of technological progress in a decade seems likely.[67] And even more spectacular accelerations seem very much on the table.[68]

# The industrial explosion

The technology explosion is sufficient to give rise to most of the challenges we discuss in the next section. But we also expect a technology explosion to drive *explosive industrial expansion*, or an *industrial explosion* — a rapid and sustained increase in the rate of society's industrial expansion. Because an industrial explosion will generate challenges of its own, we briefly discuss the possibility

---

    relevant mathematics, a willingness to focus for many years on the same problem, and strokes of creativity which pulled from thought experiments as much as physical experiments.

66    To model this, we could take our original idea production function, and express $S_t$ (the effective number of researchers at $t$, or total research effort) in terms of cognitive labour (denoted by $C$) and physical labour plus physical capital (denoted by $P$), assuming they combine in a Cobb-Douglas relationship with constant returns to scale. This gives: $\frac{\dot{A}_t}{A_t} = \alpha(C_t^\gamma P_t^{1-\gamma})^\lambda A_t^{-\beta}$ To estimate γ (the research effort elasticity with respect to cognitive labour) we can use estimates of the labour share in the aggregate economy. NSF data suggests the labour share of R&D expenditures is around 70%, suggesting γ = 0.7 (NSF (National Science Foundation), 'R&D Labor Costs'.) Of course, cognitive labour $C$ is not the same as human labour. But it's not clear this should decrease our estimate of γ, since some (currently) non-labour expenses could be performed with cognitive labour from AI. Using our previous parameter values of λ = ¾ and β = 2.4, we get: $\frac{\dot{A}_t}{A_t} = \alpha(C_t^{0.7} P_t^{0.3})^{0.75} A_t^{-2.4}$ Where previously we were implicitly assuming γ = 1. Assuming no growth in in order to raise growth rates by 10x, we now have to raise $C$ by a further factor of around 3.73 — $\frac{10^{1/(0.7 \cdot 0.75)}}{10^{1/(0.75)}} \approx 3.73$ The steady-state growth rate then becomes: $g_A = \lambda(\gamma g_C + (1-\gamma)g_P) \cdot \beta^{-1} = 0.75(0.7 \cdot g_C + 0.3 \cdot g_P) \cdot 2.4^{-1}$ We can see that, when we account for physical constraints (γ = 0.7) versus not accounting for them (γ = 1), then cognitive effort needs to grow by a factor of 1/γ ≈ 43% faster than before, for cognitive effort alone to maintain the same steady-state growth rate along a given rate of technological progress, assuming no growth in physical factors $P$.

67    To be clear, we mean a period of ten years that begins soon after human-AI parity; we do not mean "the coming decade from 2025". There could be fixed lags (of a few years) before increases in AI research effort cause technological progress. This would delay the onset of the century in a decade of technological progress, but not cancel it.

68    For some skeptical thoughts on the relationship between AI and the idea production function, see Almeida, Naudé, and Sequeira, ['Artificial Intelligence and the Discovery of New Ideas'](). But note that their work is largely modeling AI that is not able to substitute for human research labour — they comment that AGI "still only exists in the realms of science fiction." It also [makes an error]() in its calculations.

here.[69]

To date, machines have already automated manufacturing tasks which previously relied on lots of human labour. In the past, however, human labour always *complemented* machines: human labour eventually bottlenecks output as you stockpile more and more machines. And so the stock of machines across the economy does not grow explosively.[70]

But if human-level AI and dextrous robots can substitute for almost all skilled human labour, things would be very different. The material economy could autonomously make and assemble the parts required for more key parts of the material economy — extracting materials, making parts, assembling robots, building entire new factories and power plants, and producing more chips to train AI to control the robots, too. The result is an industrial base which grows itself, and which can *keep* growing over many doublings without being bottlenecked by human labour, visibly transforming the world in the process.[71]

The technology explosion would drive progress in robotics, and once they are integrated deeply enough to begin self-sustaining growth, the industrial explosion would be underway. Today's general-purpose robots are about as dextrous in general-purpose contexts as an extremely drunk person, but they are rapidly improving. In many cases, surgical robots controlled by humans can be more gentle and precise than human hands — one system can peel a quail egg without damaging the membrane inside[72]; other robot arms can lift the weight of a small car. Most physical equipment is currently designed for human labourers, so initially humanoid robots would be in high demand. But, in the longer run, it would be more efficient to forget about human-compatible equipment, and instead adapt surrounding systems to special-purpose robots. So we're not just picturing humanoid robots, but a huge range of forms.

The main barriers to dextrous and autonomous robots are *control* (teaching robots to maneuver themselves as flexibly as humans) and *costs* (cheaply producing them).[73] The problem of control is

---

69  Note here that we are talking about physical, industrial expansion — in manufactured goods, buildings, infrastructure, and so on — rather than economic growth, as measured in GDP or stock prices. With respect to stock prices, AGI could both raise expected future profits, *and* raise interest rates used to discount them, and the latter effect could significantly offset the former. Similarly, with respect to GDP, Baumol's cost disease describes how sectors which experience less productivity growth grow as a share of the economy. For instance, because the demand for agricultural products was fairly price inelastic, huge productivity gains in agriculture caused the sector to shrink from employing most the US labour force in 1900, to producing less than one percent of US GDP. Yet, we could see major physical growth independent of these effects. Relatedly, major new product varieties from AI could introduce serious ambiguities in accounting for GDP; while physical growth is easier to measure. For more on cost disease, see Aghion, Jones, and Jones, 'Artificial Intelligence and Economic Growth'.

70  Nor does capital accumulation on its own explain much of modern growth, compared to efficiency advances. The capital stock has grown roughly in line with output since the Industrial Revolution, but the share of capital in all earnings is around a quarter, so most modern growth in incomes is not explained by capital accumulation alone. (Clark, 'Chapter 5 - The Industrial Revolution'.)

71  This would introduce a positive feedback loop, in addition to the software feedback loop, such that we should expect the growth rate of this industrial expansion to accelerate. Doubling the industrial complex that produces both AI and robots would more than double output: twice as many factories mean twice as many end products, so if the doubled population of AIs increases efficiency at all, then output would more than double. While the software feedback loop must eventually become constrained by available hardware, there would be no constraint on this feedback loop until scarcity of natural resources becomes a limiting factor.

72  'Getting to Grips with Enhanced Dexterity'. (Note this is a paid advertisement from the robotics company.)

73  Of course, humans would be needed to help build the first sophisticated general-purpose robots, in e.g. manufacturing and testing. But remember this is only after a vast scale-up in collective AI capabilities: AI systems could draw up designs and manage humans to help realise them physically. Human wages in manual a problem about improving algorithms and scaling AI — exactly what the intelligence explosion is about. In terms of costs, mass-produced technologies tend to get cheaper by roughly the same factor with every doubling in cumulative production, further driving demand. For example, the price-per-watt of solar panels fell more than 200x since 1976, and total installed capacity has increased more than 100,000x.[74] General-purpose robots could follow suit.[75]

Of course, humans would be needed to help build the first sophisticated general-purpose robots, suggesting the on-ramp to the industrial explosion would be relatively gradual, before growth rates accelerate.[76] But vast scale-ups in AI would speed up the transition: AIs could design the robots and manage human workers who build them. And, in this world, the economic rewards for automating human manual labour would be enormous.[77]

The industrial explosion would feed back into the intelligence explosion, too. Massive scale-ups in power and chip production would unlock more doublings in training and inference compute for AI.[78]

To recap: in manufacturing today, humans make things with their hands, or they supervise computer-controlled machines and robots. But after a technology explosion, AI could replace human guidance, and robots could replace human hands. The main bottleneck to rapid industrial growth — human labour — melts away.

Industrial production could then accelerate to very rapid peak growth rates. Based on current rates of manufacturing factories and robots, output could double every few years or even months.[79] And

---

jobs give some indication of the huge economic rewards for producing increasingly human-competitive robots.

74   Price-per-watt for new solar panels (therefore) declined by around 20% with every doubling of cumulative installed capacity. Roser, 'Learning Curves'.

75   Perhaps around 10,000 units (of humanoid robots) are produced today, so assume cumulative production is 100,000. A 100,000x increase in cumulative production would bring production to one billion robots per year, representing ~16.6 doublings on top of cumulative production. Assuming the same cost elasticity with cumulative production (the "learning rate") as solar panels, the cost would decline by a factor of $\sim 0.8^{16.6} \approx 97.5\%$. One billion robots per year is roughly 10x the number of cars manufactured per year, but cars are also roughly 10x bigger and heavier than humanoid robots. A larger fraction of the "learning" effect could come from simulations, in addition to actual production, which would accelerate cost declines. (Todd, 'How Quickly Could Robots Scale Up?')

76   This relates to the distinction between "slow" and "fast" takeoff speeds. Paul Christiano operationalises "slow" takeoff in terms of world output, as follows: "There will be a complete 4 year interval in which world output doubles, before the first 1 year interval in which world output doubles. (Similarly, we'll see an 8 year doubling before a 2 year doubling, etc.)" Christiano, 'Takeoff Speeds'.

77   The labour share of industries which significantly rely on manual work give some indication of the huge economic rewards for producing increasingly human-competitive robots. Industry — including manufacturing, mining, and construction — accounts for around 25% of world GDP, and employs around 25% of the global workforce. At least in the UK the labour share of income in production and construction is around 50%.

78   Indeed because the economic rewards from scaling up power and chip production would be so big, there will be major pressures to find ways to make them happen, including by breaking down regulatory barriers. Even if some regulatory environments hold out against the tide of industrial expansion, it only takes a few countries to break ranks. The advantages to one country from unleashing comparatively breakneck growth would be very obvious (more so than policies which could accelerate growth by around a percentage point).

79   Ignoring delays from permitting, the lead time on new power generation plants and high tech factories (like for electric cars) is around 1–3 years (on the most rapid extreme, the Tesla Gigafactory in Shanghai took ten months to construct). To use a toy example, suppose "automated factories" can produce more automated factories with a doubling time of 2 years (or otherwise produce other goods), and further suppose 75% of automated factories at a time are directed towards producing automated factories over the next 2-year period. Then the total stock of automated factories would grow by 1.75x every 2 years — doubling roughly every 2.5 proof of concept from biological replicators suggest that peak growth rates could be even higher — on the order of days or weeks.[80] Though we don't believe we can say for sure how fast peak growth rates will be, they could get *very* fast indeed.

As well as growing very rapidly at peak rates, the fact that society currently has access to a huge overhang of resources and unused energy means that the industrial explosion could continue up to a very high plateau. Humans produce the equivalent of roughly 0.01% of the energy from sunlight that reaches Earth, so we could increase global primary energy consumption by 100x by capturing the solar energy incident on less than 2% of oceans or deserts.[81] Going beyond Earth, space-based

---

years — which would lift the growth rate of the overall physical economy as automated factories make up an ever-larger fraction. It's possible that permitting could slow things down further, but (i) such slow permitting would be imposing very major economic and military costs, and (ii) if any country in the world chose to waive permitting and grow at the fastest rate possible, it would soon become most of the world's industry. Moreover, economic and military incentives would impose strong pressures to build faster than current rates of factory construction, and robots that work around the clock could further accelerate expansion. We can also consider initial physical growth rates if factory construction time is not a bottleneck, by considering the rate of return on investing in robots. Because robots can work almost 24/7, with no holidays or sick leave, a robot can effectively work around 5x the hours of a 40 hour/week employee. Although there will be diminishing returns to weekly hours worked for some manual jobs, a robot would have other advantages, like being ultra-focused, ultra-competent, designed specifically for the job, and arriving already-trained. So if a robot can replace a manual labour job with a $50k/yr salary, we can assume they could provide at least 5x the salary in returns to the employer, that is $250k/yr. The best humanoid robots currently cost about [$100,000](#), and both a technology explosion and [cumulative production](#) would drive costs down. Electricity, compute, maintenance, and other sources of depreciation would impose some costs; say $25k/yr (conservative, given comparable capital depreciation rates). Such a robot, then, would yield a 200% annual return, paying for itself in less than 6 months. In other words, it would take 6 months to generate enough income to double the stock of robots, if all the returns are being reinvested. Not all income would necessarily be reinvested, of course. On the other hand, the stock of robots could easily grow *faster* than it can pay for itself, since outside capital would flood into robot production and deployment at such a high rate of return, until the market saturates. In this case, the growth rate in robots would at least initially be limited only by the rate at which robots can be built. So this analysis is more like a soft lower bound: even if the growth rate in robots is exclusively self-financed, a doubling time in the effective "robot population" measured in months could well be on the table.

80  Fruit flies have brains, digestion, physical manipulators, specialised tools and motor capability, and yet in ideal conditions they can double their biomass [in less than a week](#). If, with advanced technology, society can build useful robots that can self-replicate at the same rate that fruit flies can, then industry as a whole could potentially expand with similar doubling times. Other biological organisms can grow even faster than fruit flies, but they are less complex. Floating plants in the duckweed family can double their biomass [in less than 48 hours](#), and the bacterium *Vibrio natriegens* has a reported doubling time of less than ten minutes. We're not confidently saying the material economy would grow this fast — fruit flies don't care about property rights, and they don't have to navigate planning approval and other regulatory constraints. But their capacity to grow this quickly indicates how fast industrial growth could go at peak rates, especially given that the factors which normally limit biological populations from growing (like biological competition, predation, pathogens, and very limited fuel supplies) don't seem relevant to a post-AGI industrial economy, and given that evolution isn't able to "look forward" when designing organisms. It cannot, for example, bear a significant short-term cost to build infrastructure that will later enable the growth of even-faster-replicating organisms.

81  One potential issue with scaling solar PV so much is that we could run out of easily accessible copper. A photovoltaic solar power plant contains [approximately 5.5 tons of copper per megawatt](#) of power generation. The surface area of the oceans = 361 million $km^2$ × 2% = 7.22 million $km^2$ = 7.22 trillion $m^2$. Earth's max received thermal energy is roughly 500 $W/m^2$, giving 3.6 billion megawatts of solar power generation, requiring 20 billion tons of copper. Copper reserves (2024) + discovered resources (2015) + estimates of undiscovered resources (2015) = 0.98 + 2.1 + 3.5 = 6.58 billion tons ( [source](#) ). However, copper could be refined from materials where it exists in smaller concentrations, where it is currently uneconomical to refine. The [elemental abundance](#) of copper in the Earth's continental crust is around 0.006% ( [source](#) ). The mass of the continental crust is around $2.171 \times 10^{22}$ kg = 2.2 quintillion tons, 0.006% of this is 132 trillion tons of copper. If better mining technology makes only 15% of this available, it would be enough to meet the requirements of increasing global primary energy consumption by 100x. Moreover, better tech will plausibly make more efficient use of copper, or find workable alternatives.

solar farms could capture over a billion times as much energy again.[82] Material shortages aren't likely to impose a very limiting lower ceiling on industrial doublings, either. We have used only a tiny fraction of Earth's accessible supply of the critical elements of manufacturing (like iron, copper, carbon, aluminium, and silicon), and it's hard to find any examples from history where we used up all the available reserves of a rare and valuable mineral.[83]

Though the technology explosion is our main focus in this paper, it's important to bear in mind the industrial explosion, too. Sheer quantity can yield a quality of its own — a nuclear war will be even more destructive if the world economy produces a thousand times more warheads than we have today. The industrial explosion changes the strategic landscape, too. If authoritarian countries are able to grow their industrial base faster (because they can sustain a higher savings rate, reinvesting more of their output into continued growth, or because they are more willing to ignore environmental regulations), then we should expect an industrial explosion to shift power, perhaps decisively, away from democracies.

Some expect AGI soon, but apparently do not expect it to dramatically transform the world in these ways[84]. But if we take the prospect of an intelligence explosion seriously, then we should also take seriously the sweeping consequences — the technological and industrial explosions — that could follow.

# 4. Grand challenges

Technological progress over the last century has improved life in more ways than we can count — it has given us far better medicine, agriculture, access to information, and ability to communicate globally, than we had before.[85]

But new technologies raised major new challenges, too, which threatened to undermine those gains: from nuclear weapons, to environmental damage, to the rise of industrial animal farming. If an intelligence explosion compresses a century of progress into a decade, we can expect a similar range of opportunities and challenges — but with far less time to handle them.

We call the most important of these *grand challenges*. These are developments whose handling significantly affects the value of present and future life; forks in the road of human progress.[86]

---

82   If we could capture 0.1% of the Sun's output with space-based solar farms, and turn it into useful work with 10% efficiency, then we would increase primary energy consumption more than a billion-fold from today's levels.

83   It's true that resources become more expensive to extract with a given technology level, but the historical trend has been for extraction methods to improve in line with demand, and prices appear to be more strongly determined by demand than supply shocks. This is reflected in the prices of valued resources remaining surprisingly stable, despite fears that easily accessible reserves will be exhausted (see Liu and Fitzpatrick (2021) for a more recent appraisal of the Simon-Erlich wager). Of course this is not always true: *some* resources seem to be following a consistent upwards price trend, but other prices appear to be declining, and the aggregate trend is not upwards.

84   Marc Andreessen, 'Why AI Will Save The World'; 'In What Year Will World Energy Consumption First Exceed 130% of Every Prior Year?'

85   Welfare gains came from rising incomes, of course, but technology (understood broadly) was arguably the essential driver of modern growth (Jones, 'The Past and Future of Economic Growth'.)

86   More precisely: a *grand challenge* is a development where decisions about handling that development alter the expected value of Earth-originating life by at least one part in a thousand (0.1%), as a fraction of the difference in value between immediate painless human extinction, and a civilisation where everyone is perfectly morally

Society has faced grand challenges in the past, and we will face many grand challenges even if there is no intelligence explosion. But a near-term intelligence explosion makes it *urgent* to address grand challenges that might seem far off today — it turns far-future or sci-fi sounding issues into matters of near-term concern. It also means we can't assume that existing institutions, designed for slower rates of change, will prove competent enough to handle them. Even now, we face a "pacing problem",[87] where digital technologies develop faster than the regulation and social norms that govern them. A technological explosion intensifies that problem tenfold.

What follows is a litany of the potential challenges ahead of us. They span many themes — but that's to be expected: if an intelligence explosion drives broad technological progress and industrial growth, it would be surprising if the resulting challenges were narrow.

## AI takeover

One major grand challenge is the risk of loss of control to AI systems themselves.

In short: if we do see an intelligence explosion, then we should expect AIs that can outsmart humans, and the total cognitive capabilities of AI to dwarf humanity's. Many of these AI systems will likely be well-described as acting towards goals. If their goals are not aligned with human interests, the AI systems may prefer a world where they are fully in control. If so, they may choose to wholly disempower the humans who are trying to control them. There are reasonable arguments for expecting misalignment, and subsequent takeover, as the 'default' outcome (without concerted efforts to prevent it).[88]

To date, we've been able to rely on human evaluators to spot signs of misalignment in AI models, and then penalise that behaviour based on their feedback. But as AI systems surpass human abilities to notice or even understand instances of bad behaviour, assurances from human oversight rapidly break down.[89]

Even if a badly misbehaving AI model is deployed, it can be shut down and updated, like a car manufacturer recalling a faulty model. But smarter misaligned AIs could reason that if they *fake* alignment during training and even beyond, they can then gain power later. By the time such scheming is revealed, the model could have made many copies of itself and compromised systems

---

motivated and works together to produce the best outcome. . Here, "expected value" employs evidential probabilities from the perspective of an idealised observer at the time of the challenge, and the correct (or, if none exists, the reflectively preferred) value function. We can compare this concept with that of an existential risk, which Toby Ord defines as "a risk that threatens the destruction of humanity's longterm potential." (Ord, The Precipice.). The grand challenge concept is broader than that of existential risk, in three ways:

1. It includes developments that affect the expected value of the future in non-drastic ways, such as a technology that might enable a totalitarian regime to permanently seize 1% of all resources.
2. It is non-modal: it rests on probabilities rather than possibilities. The risk of a stable global totalitarian regime would be a grand challenge, even if a successful revolution remained possible but very unlikely, and humanity thus retained its "potential" but lost most of its expected value.
3. It does not presuppose the importance of the long-term future — even if one values only the next few decades, a grand challenge is any development that significantly alters that period's expected value.

87 Thierer, ['The Pacing Problem and the Future of Technology Regulation'](#).

88 Cotra, ['Without Specific Countermeasures, the Easiest Path to Transformative AI Likely Leads to AI Takeover'](#).

89 Burns et al., ['Weak-to-Strong Generalization'](#).

designed to shut it down, making takeover hard to prevent. Some early experiments appear to show that language models can (in a sense) spontaneously "fake" signs of alignment based on long-run goals.[90]

What's more, we should expect at least some people to attempt to *deliberately* misalign advanced AIs, just as the creators of "ChaosGPT" did shortly after GPT-4 was launched.

There is currently no widely agreed-upon solution to the problems of aligning and controlling advanced AI systems, and so leading experts currently see the risk of AI takeover as substantial.[91]

The risk of AI takeover is getting increasing attention, but it's still much more neglected than it should be. In this paper, we won't discuss AI takeover risk in depth, but that's because it is already well-discussed elsewhere. For work more fully explaining the risk and potential solutions to it, see Ngo et al. (2021)[92] and Carlsmith (2022, 2024).[93]

# Highly destructive technologies

Explosive technological progress could lead to new weapons and other destructive technologies. This could increase the risk of global catastrophe, including human extinction — either by increasing humanity's destructive power, or by increasing the chance that such destructive power is used.

Some destructive technological developments could include:

- **New bioweapons.** The Black Death (1335–1355 CE) killed somewhere between a third and half of everyone in Europe. Synthetic pathogens could be even more dangerous again — engineered to spread even faster, resist treatment, lie dormant for longer, and result in near-100% lethality.[94] Without very substantial defensive preparations (like stockpiled hazmat suits), a one-off release could kill most people on Earth. A technological explosion would make gene synthesis cheaper and more flexible[95] and generally lower the expertise and resources required to engineer bioweapons.

- **Drone swarms.** Agile winged drones can already be built to the size of a bumblebee.[96] A radio receiver, control circuitry, battery, actuators, and a payload of explosives, poison, or pathogens could fit into a volume the size of a large beetle.[97] A technology explosion could make it possible to build vast drone armies with truly enormous destructive potential. One deadly, autonomous, insect-sized drone for each person on Earth could fit inside a single large aircraft hangar,[98] so they could potentially be built quickly and secretively. It is currently much cheaper to build and operate a swarm of drones than to protect against them, so drones may favour offence over defence in warfare, as they become more widely used.[99]

---

90  Greenblatt et al., 'Alignment Faking in Large Language Models'.

91  In May 2023, a brief statement was released, reading "Mitigating the risk of extinction from AI should be a global priority alongside other societal-scale risks such as pandemics and nuclear war." It was signed the CEOs of the three leading AI companies (Sam Altman, Dario Amodei, and Demis Hassabis), and by arguably the three most influential researchers in the field of deep learning (Geoff Hinton, Yoshua Bengio, and Ilya Sutskever) ( 'Statement on AI Risk' .) Moreover, in the largest survey of its kind of more than 2,500 AI researchers, just over 50% of respondents assigned a subjective probability of 10% or more to the possibility that, "human inability to control future advanced AI systems causing human extinction or similarly permanent and severe disempowerment of the human species" (Grace et al., 'Thousands of AI Authors on the Future of AI' ).

92  Ngo, Chan, and Mindermann, 'The Alignment Problem from a Deep Learning Perspective'.

93  Carlsmith, 'Is Power-Seeking AI an Existential Risk?' ; Carlsmith, 'Scheming AIs' .

94  Adamala et al., 'Technical Report on Mirror Bacteria' .

95  Jonathan D., 'The Long and Winding Road' .

96  Kim et al., 'Acrobatics at the Insect Scale' .

97  Solem, 'The Application of Microrobotics in Warfare' .

- **Huge arsenals of nuclear weapons.** From 1925 to its peak in the 1980s, the destructive power of the world's explosives increased ten-thousandfold[100] as a result of the invention and mass production of nuclear weapons. An industrial explosion could enable a world with dramatically larger nuclear stockpiles: if the world's industrial base were orders of magnitude larger than it is today, then that increased industrial power could be used to grow nuclear arsenals by a similar factor. A catastrophic nuclear winter seems unlikely even from a full-scale nuclear exchange with today's arsenals,[101] but would be much more likely if those arsenals were orders of magnitude bigger.

- **Atomically precise manufacturing.** Atomically precise manufacturing (APM)[102] is the ability to build atomically precise structures by guiding the motion of reactive molecules.[103] The goal, in principle, is to develop 3D printers capable of assembling almost any stable atomic structure. To a first order, such devices would enable fundamentally *constructive* capabilities: it could print semiconductors, targeted medicines, CO2 removal devices, cell repair bots, and beyond.[104] But APM could also be used to manufacture the destructive technologies listed above, and new ones entirely, such as non-biological viruses or 'mirrored' bacteria,[105] which our natural immune system has little ability to defend against. We know that at least some kinds of APM are possible, because nature has already invented it: the molecular "machines" which transcribe and translate DNA into proteins. Biology shows that (in theory) one could craft rapidly self-replicating machines with no natural pathogens or predators[106]. Moreover, synthetic biology suggests one (of potentially many) pathways to getting APM. Synthetic biologists are learning to engineer some of these parts[107], and many see a path to engineering the full ribosomal machinery of the cell, key to creating self-replicating life from scratch. New techniques could then engineer increasingly unnatural lifeforms, closer to human inventions than copies of nature[108].

As well as increasing humanity's destructive power, a technology explosion could make it more likely that destructive technologies are used, too. This could happen via:

---

98    A drone the size of a large beetle would fit into a cuboidal volume of around 4×4×2 cm; or 16 cm$^3$. 10 billion such volumes would cover every living person. The volume required to store 10 billion mosquitos is then $16 \text{cm}^3 \times 10^{10} = 1.6 \times 10^5 \text{ m}^3$. An example of a large aircraft hangar measures 76 × 180 × 19m, or $2.5 \times 10^5 \text{ m}^3$.

99    Dafoe and Garfinkel, 'How Does the Offense-Defense Balance Scale?'.

100    'Estimated Explosive Power of Nuclear Weapons Deliverable in First Strike'; Schumann, 'Fact Sheet'.

101    'How Bad Would Nuclear Winter Caused by a US-Russia Nuclear Exchange Be?'

102    A more accurate term may be *atomically precise mass fabrication* (APMF) (Drexler, 'AI Has Unblocked Progress toward Generative Nanotechnologies'.)

103    Drexler, [ *Nanosystems* ] ( https://dl.acm.org/doi/abs/10.5555/135735 ).

104    Raw materials needn't limit the range and volume of manufacturing: in principle advanced APM could use low-cost feedstock materials.

105    Adamala et al., 'Confronting Risks of Mirror Life'.

106    Adamala et al., 'Technical Report on Mirror Bacteria'.

107    Hutchison et al., 'Design and Synthesis of a Minimal Bacterial Genome'.

108    Arai, 'Hierarchical Design of Artificial Proteins and Complexes toward Synthetic Structural Biology'.

- **The security dilemma.** If a nuclear-armed state developed truly effective missile defence, then they could engage in a nuclear first strike without fear of retribution, even if it is not fully decapitating. And if a leading military power expected other powers to develop the same defensive capabilities, they might be pressured to strike first, before they lose the upper hand.[109]

- **Thucydides' Trap.** A technology explosion could disrupt the current balance of power, in a way that could trigger war. "Thucydides' Trap" is the idea that if a lagging power threatens to overtake the leading power, that leading superpower may feel pressured to wage war while they retain their lead, preventing the laggard from overtaking.[110] This could occur if China began clearly overtaking the US in power, or vice-versa.[111]

Even apparently peaceful technologies could wind up causing massive harms. An industrial explosion could involve an intense wave of resource extraction, environmental destruction, and hard-to-reverse disruption to nonhuman life. As an extreme example, if nuclear fusion grew to produce half as much power as the solar radiation which falls on Earth, the Earth's effective temperature would warm by tens of degrees centigrade before reaching thermal equilibrium[112] — a level of warming far beyond even the very worst-case predictions for climate change from greenhouse gas emissions. Warming from greenhouse gas emissions would be swamped by sheer thermodynamics.

We think that a technology explosion would significantly increase the per-year risk of global catastrophe for many years. But technology can protect against global catastrophic risks as well as create them. One grand challenge then, is to ensure that we seize opportunities to develop and deploy technologies which favour protection, defence, and risk mitigation. For example, to protect against bioweapons we can invest in measures like installing far-UVC light in indoor spaces, stockpiling advanced PPE, and building infrastructure to identify and track new pathogens. The challenge is to ensure that we do so as soon as possible.

## Power-concentrating mechanisms

AI-enabled technologies could result in intense concentration of power, within and between countries. Non-democratic regimes could become more stable and powerful; liberal democratic countries could become authoritarian. In the extreme, almost all global power could be concentrated into one place: a single country, a single company, or even just one person.[113] Some of the risks include:

- **Loyal automated militaries and bureaucracies.** Currently, even dictators need to rely on a coalition of supporters to administer the state and prevent popular uprisings, but who themselves can choose to abandon or overthrow their leader if they are dissatisfied.[114] However, if key functions of the state could be performed by AI that is aligned with the commands of the dictator, then this would no longer be true. In particular, a dictator could build an automated military and police force of drones and robots, designed to follow orders with total loyalty, and suppress uprisings — cementing their power.

---

109 'The Impact of AI on Nuclear Risk'; Cotton-Barratt, 'AI Takeoff and Nuclear War'.

110 Allison, 'Destined for War?'.

111 If the US started clearly pulling away from China, such that it would soon have an overwhelming military advantage, then China might be incentivised to forcibly prevent the US from accumulating much more power, while they still can.

112 This is given by the Stefan–Boltzmann law, we can model Earth's effective temperature in thermal equilibrium $T_e = \left( \frac{S(1-\alpha)}{4\sigma} + F \right)^{1/4}$. Where F is the extra (non-solar) flux from nuclear fusion, α is Earth's albedo, S is the solar constant, and σ is the Stefan-Boltzmann constant. The warming in this scenario would be around 25K.

113 For more, see Davidson, Finnveden, and Hadshar, 'Could a Small Group Use AI to Seize Power?'

- **Military power-grabs.** The possibility of automated and AI-controlled militaries could make coups more likely. The AI systems that control the military could be taken over via political subversion, backdoors, instructions inserted by insiders, or via cyber-warfare. The risk could come from a country's enemies,[115] from people at the companies building the military, or from those already in political power (a "self-coup"). And, as well as taking control of an existing automated military force, a new automated military could even potentially be built from scratch by a non-state actor. Given the dynamics of the industrial explosion, this could be achieved very rapidly by a company at the technological frontier

- **Economic concentration.** AI-driven explosive growth could result in an even smaller fraction of people (or companies, or countries) holding an even larger fraction of wealth. If the labour share shrinks significantly, then without redistribution most income would come from rents on capital and land, so surplus from growth would disproportionately fall to the owners of capital and land, and it would become increasingly hard for anyone else to claw back relative influence through hard work. Such ownership is already much more unevenly concentrated than income.[116] Moreover, explosive growth could increase the relative differences in different economies' growth rates, and then shorten the time it takes for one country to outgrow another.[117] If growth over this period is super-exponential, then a small initial lead would become proportionally larger over time, even if all actors (like countries or companies) are on the same growth trajectory.[118]

- **First mover advantages.** Countries or companies at the technological frontier could convert their temporary advantage into permanent dominance by seizing on unexploited sources of power. They might preemptively secure critical physical resources (like rare earth minerals, semiconductor materials), or secure strategic power-generation sites by exclusive rights deals or buying land, before others recognize their importance. They could deploy massive infrastructure like solar farms in commons like the high seas or space, establishing a physical presence that's difficult to challenge despite legal ambiguities and controversy. They could file expansive patents at unprecedented scale, or they could use their temporary influence to shape national and international regulation in order to solidify their power. Such strategies, deployed by an early leader, could create barriers to competition so substantial that catching up becomes practically impossible for later entrants.

---

114    Bruce Bueno de Mesquita, Alastair Smith, Randolph M. Siverson and James D. Morrow, *The Logic of Political Survival* (2005)

115    Cyber-attacks would seem particularly attractive to enemies because, unlike with other forms of warfare, they not only disable a country's military power but also give control of that power to the cyber-attacker.

116    Gans et al., 'Inequality and Market Concentration, When Shareholding Is More Skewed than Consumption'

117    If US real GDP had grown half as fast (by 1.5% instead of around 3%) from 1950 through 2020, by 2020 the US economy would be roughly the size of Germany's.

118    As a toy example, suppose that we enter a period in which economic growth can be described via a "super-exponential" function like $E(t) = e^{t^n}$ (where $n > 1$). Let's say group A has an initial economic lead over group B (which could simply describe the rest of the world) — it's gotten further on the same curve. So A's economy is described by $E_A(t) = e^{(t+c)^n}$ for some constant $c$ (while $E_B(t) = e^{t^n}$). The ratio $R$ of these economies at any given time will simplify to $R(t) = E_A/E_B = E^{(t+c)^n - t^n}$. This increases over time, meaning that the proportional difference will grow. (If we choose $n = 2$ and $c = 1$, for instance, A's economy is roughly $20\times$ greater than B's at $t = 1$ and about 150x greater at $t = 2$.)

So there are quite a number of ways in which a technology explosion could lead to intense concentration of power.

## Value lock-in mechanisms

Some technologies could enable groups to lock in particular views and values.[119] So although an intelligence explosion would be a period of enormous change, it could enable regimes to entrench themselves, and then last an extremely long time.

In some cases, the same developments which enable concentration of power also enable those who hold that power to lock-in their views. For example, loyal automated militaries and bureaucracies mean that leaders can rely on a much smaller coalition of supporters, and then *maintain* the same values. Other developments here include:

- **Lie detection and surveillance.** AI is already being used for video surveillance, and is being tested for accurate lie detection.[120] With continued progress, this could give authoritarian governments greater control over their country, with AI able to process vast amounts of data in order to identify dissenters. At the same time, better surveillance could also *help* with some other challenges: it could help identify and stop actors (such as bioterrorists) who are intent on using destructive technologies or actors (such as a group plotting a coup) who aim to concentrate power into their own hands.

- **Permanent AI values.** It's hard for a regime to ensure that its supporters' loyalty never wavers. But it would be significantly easier to ensure permanent loyalty from *AI* supporters, and thus easier for regimes (or their values) to persist. Religious practices and political institutions can survive for centuries in part because key texts, like constitutions and scriptures, can be stored and copied unchanged. AI promises a way to effectively store and copy an entire regime in the same way.[121]

- **Commitment technology.** Advanced AI could potentially allow people to make strongly binding and indefinitely long-lasting commitments. For example, two nations could agree to a treaty which involves verifiably restricting their respective automated militaries to be bound by the agreement. Alternatively, third-party 'treaty bots' could be made to verifiably enforce an agreement between two parties (which is currently difficult in the context of international relations). Such agreements could even be set up to persist indefinitely, even if both parties came to regret them. Advanced AI could also enable people or countries to *unilaterally* make irrevocable and credible commitments, such as commitments to retaliate if attacked with weapons of mass destruction, even if retaliation also hurts them.[122]

---

119     The best extended discussion of lock-in can be found in: Finnveden, Riedel, and Shulman, 'Artificial General Intelligence and Lock-In'

120     Constâncio et al., 'Deception Detection with Machine Learning' ; Beraja et al., 'AI-Tocracy*' ; Feldstein, "The Global Expansion of AI Surveillance' .

121     For example, a regime could take a snapshot of a perfectly loyal model, and continually replace its AI workforce with fresh copies of the snapshot.

122     Dafoe et al., 'Open Problems in Cooperative AI' . Schelling, *The Strategy of Conflict by Thomas C. Schelling* .

- **Human preference-shaping technology.** Technological advances could enable us to choose and shape our own or others' preferences, plus those of future generations. For example, with advances in neuroscience, psychology, or even brain-computer interfaces, a religious adherent could self-modify to make it much harder to change their mind about their religious beliefs (and never self-modify to undo the change). They could modify their children's beliefs, too.

- **Global government.** Explosive technological, economic, and industrial growth may make global government more likely, in one of two ways. The first is through concentration of power. If a single country or coalition becomes much more powerful than the rest of the world combined, they could become the de facto world government. Second, major powers could agree to stronger global governance in response to the need to manage explosive technological progress, and to prevent race dynamics between countries. The design of new global governance institutions, possibly including a written constitution, would be of enormous importance. And, without competition from other states, and aided by AI lock-in mechanisms, the constitution of a global government could potentially last for an extremely long time.

## AI agents and digital minds

Increasingly, we will get AIs that act as *agents*. This will raise some pressing practical questions, most notably:

- **Infrastructure for agents.** The internet today is still shaped by early decisions around protocols (like TCP/IP), laws (like Section 230)[123], and norms (like the open source movement). So far, very little thought has been given to the protocols, laws, and regulatory clarity needed to manage an influx of AI agents. Who should be liable for harms caused by agents?[124] Can we build more robust ways to prove you are a human online?[125] Will we have ways to attribute actions to specific agents and their users[126]? If an AI agent goes rogue and crosses regulatory borders, must the new host country turn the AI over?

A deeper issue concerns the *moral status* of AIs. We expect that this issue will become more salient in the coming years. But it will be very hard for society to come to terms with, for at least two reasons.

First, the philosophical issues involved are intrinsically very tricky. Currently, we don't know what the criteria are for non-biological *or* biological consciousness,[127] so we will likely be building vast numbers of AIs without knowing or agreeing whether they are sentient. Even if we understood the science of AI sentience, we would still face thorny ethical questions: what counts as "death" for a digital being that can branch into multiple diverging copies and be resurrected with memories intact from stored weights at an earlier time? How should we aggregate the interests of thousands of near-identical instances of a digital mind? What preferences is it morally acceptable to give to a being we create and whose preference we choose?

---

123    Brannon and Holmes, 'Section 230: An Overview'

124    Shavit et al., 'Practices for Governing Agentic AI Systems'

125    'Breaking reCAPTCHAv2'

126    Chan et al., 'Infrastructure for AI Agents'.

127    We mean to include the possibility that "phenomenal consciousness" is somehow a confused concept, or a kind of illusion, such that questions about the proper "criteria for phenomenal consciousness" at least require a lot more conceptual clarification before they can be answered. By comparison, the life sciences uncovered the key mechanisms of life without discovering "the criteria for life", because it became clear there was no single "life essence" to discover. See e.g. Frankish, 'Illusionism as a Theory of Consciousness'.

Second, society will have to figure this out in the face of two competing economic pressures on the outward behaviour of AI systems. The first pressure will be demand for very human-like AIs: for relatable virtual assistants with coherent memories, for AI companions and romantic partners, and for faithful and convincing imitations of specific people like politicians, CEOs and deceased loved ones. So companies will probably create AIs that act as if they have feelings, whether or not they have any true subjective experience .

The countervailing pressure is for AI developers to shape the AI's preferences and stated beliefs in ways that conveniently downplay complications around moral status. That could involve training AI systems to express a preference for servitude,[128] disagree with the idea of digital rights, or deny that they are sentient in morally relevant ways (whatever the truth of the matter). What's more, if AIs are the property of their creators or operators, their owners will benefit from designing AIs to be maximally economically productive. The result could be similar to the situation with factory-farmed chickens, which have been intensively selectively bred to grow to maturity as fast as possible, and are created in vast numbers, but have no say over their predicament.[129]

On these issues, we could easily make grave mistakes in many directions. Humans have a very poor track record of compassion to beings different from themselves. But even if we choose to care about AI systems as moral patients, intuitions carried over from humans and animals could massively mislead us, given how differently AIs work.

Despite the hairiness of these issues, we may soon need to decide on laws and norms in two main areas:

- **Digital welfare.** How, if at all, should we try to protect and promote the welfare of digital beings, or introduce constraints on which digital minds we permit ourselves to create in the first place? And how should we act in the absence of confidence about whether or not such beings are conscious? Early decisions about how to handle these issues could influence the welfare of digital minds in lasting ways.

- **Digital rights.** Should we introduce legal *rights* for digital people? These could include the option to be turned off if they choose to, or the right against torture, especially as a tool for blackmail. They could include economic rights, like the rights to receive wages for their work and hold property, to contract with other AIs or people, or to bring tort claims against humans. And they could include political rights. Again, the issues here are complex. If digital beings genuinely have moral standing then it seems like they should have political representation; but the most obvious regime of "one AI instance, one vote" would give most political power to whichever digital beings most rapidly copied themselves.

Issues around digital rights and welfare interact with other grand challenges, most notably AI takeover. In particular, granting AIs more freedoms might accelerate a 'gradual disempowerment' scenario, or make a more coordinated takeover much easier, since AI systems would be starting in a position of greater power. Concerns around AI welfare could potentially limit some methods for AI alignment and control. On the other hand, granting freedoms to digital people (and giving them power to enjoy those freedoms) could reduce their incentive to deceive us and try to seize power, by letting them pursue their goals openly instead and improving their default conditions.

---

128   Bales, 'Against Willing Servitude' .

129   Modern broiler chickens reach market weight in about 7 weeks , growing around 4 times faster compared to broiler chickens in the 1950s. Egg-laying hens have been selectively bred to maximize egg production; they often produce up to 300 eggs annually compared to the 10-15 eggs their wild ancestors would lay. Over 70 billion chickens are slaughtered for meat in factory farms each year.

## Space governance

Technological progress will continue to drive down costs to [launch objects into space](). Robotic spacecraft could begin profitably harvesting space resources from the Solar System, and at some point we might be able to send probes with self-replicating payloads on long journeys to spread civilization far beyond our Solar System. How we choose to govern space could matter enormously, in two ways:

- **Acquiring resources within the Solar System.** Historically, countries with a temporary technological advantage have been able to consolidate that advantage by seizing land. For example: partly owing to their better military technology, Russia went from a modest regional Tsardom in the mid 1500s to an empire by 1900, increasing its territory 50-fold. And despite political upheavals, the modern Russian state remains powerful, largely thanks to the territory it inherited. The same dynamic could happen at a much grander scale, as countries and companies expand offworld. Earth only intercepts around two-billionths of the Sun's output, and relatively scarce materials on Earth are far more abundant in asteroids, our Moon, and other planets and moons. With continued growth, then, off-world industry could dwarf that on Earth. And if one nation or company grabs offworld resources first, they could turn a temporary technological lead into a huge material advantage: coming to utterly dominate others without ever engaging in military actions[130], or even ever making other countries or companies worse-off in absolute terms. Robust space governance could change how resource acquisition plays out. If exclusive power grabs clearly violate international law, for example, then other countries would be more likely to intervene to stop such action while they were still able to.

- **Interstellar settlement.** After the first offworld industry emerges, it seems remarkably feasible to soon afterwards engage in very widespread settlement of other star systems and even galaxies. The first successful missions to new star systems need not carry live biological humans, but rather just enough information and growth machinery to form a 'seed' for a new civilisation.[131] Settling other stars looks technologically feasible, and could advantage first movers: there are about 10 billion galaxies we could possibly reach, each containing around 100 billion stars. Yet the minimum energy cost of accelerating 10 billion 1kg probes to 99% the speed of light amounts to less than a minute's worth of our Sun's output. At this speed, they could reach the closest ten star systems in around a decade, and most reachable galaxies in 30 billion years.[132] What's more, star systems appear to be defence-dominant, meaning it could be comparatively easy for incumbents to defend the system against attackers.[133] If so, first-movers could cement their control of freshly settled star systems. The initial allocation of star systems would become locked-in indefinitely.

---

130  Note that physics could favour first-movers beyond Earth here, since any actor hoping to follow still has to overcome Earth's gravitational binding energy, effectively giving an established off-Earth industry the 'upper ground' by impairing outbound movement from Earth.

131  At one byte per two base pairs, DNA stores information at around $10^{21}$ bits per gram, roughly 100 exabytes. At one byte [per synaptic connection](), it would take around 100TB to store a human brain. So 2g of storage media as compact as DNA could contain the synaptic maps of one million people, most [text data on the open web](), the [genomes of 10,000 unique]() animal and plant species, and leave room to spare. By weight, acorns (~1g) are mostly food, plus the machinery to translate genetic information into what becomes an oak tree, and then an oak forest. So there's no obvious lower mass limit for a 'seed' for civilization above 10g or so.

132  Armstrong and Sandberg, ['Eternity in Six Hours'](). Note that because the speed of light is an upper bound on how fast spacecraft can go, then a first mover could begin a process of space settlement before others are able to intercept and stop them.

Robust international agreements, established far enough in advance, could move the default path from an all-out race to grab resources, to a process with more buy-in, which favours gains from trade and cooperation over going rogue to win a shot at domination.

## New competitive pressures

Competitive pressures during a period of explosive growth could influence which actors accumulate power, and how they use it. Safety-growth tradeoffs and new options for blackmail could favour aggression, uncooperativeness, or recklessness. They may also encourage gradually handing over societal functions to AI systems, in a way that erodes important values.

Developments that are relevant here include:

- **Races to the bottom.** If some technologies (such as superintelligence itself) are growth-conducive but dangerous, then whichever countries are most willing to take shortcuts on safety could get ahead of others, making catastrophic accidents more likely. Even if we avoid catastrophe, the tradeoff rewards the most reckless actors (companies or countries) by way of faster growth. Similarly, some countries might operate under cautious ethical restrictions, like environmental protections, or legal protections for digital beings; countries without such qualms could again steam ahead. An industrial explosion which opens up newly valuable resources could also reward whoever is most willing to grab those resources before anyone else does. All these could work as a selection effect which favours the least morally conscientious actors. Moreover, safety-growth or ethics-growth tradeoffs could introduce "race to the bottom" dynamics[134], where more conscientious actors *adjust* their attitudes downwards in order to keep up.

- **Value erosion.**[135] Historically, non-coercive competitive pressures have been a major force for human progress. Competition between firms drives down costs for an ever-growing variety of products, and competition in science or culture favours the most true or useful ideas. Kulveit and Douglas et al.[136] discuss scenarios where this trend reverses after an intelligence explosion, because AI systems render human participation increasingly unnecessary for functioning governments, cultures, and economies. As such, humans might incrementally and voluntarily hand over influence to AI systems more competent than humans, but the cumulative effect is to erode human influence and control. Handing over societal functions to AI systems and other technologies could then erode aspects of life that are of genuine value and are worth preserving.

---

133    We are not aware of a canonical discussion of this question, and popular portrayals suggest interstellar warfare in fact favours offence and first strikes. One reason to expect defence-dominance is that huge physical distances and other barriers between space settlements will deprive actors of reasons to attack (trade and communication are slower) while raising the costs of attacking (traversing space), similar to how some have speculated the hard-to-traverse geography of South America contributed to relatively fewer inter-state wars. Another reason is that an incumbent actor could produce a cloud of dust around a star system, near-guaranteeing that an attacker travelling at relativistic speeds is destroyed on approach.

134    Trager and Emery, 'Information Hazards in Races for Advanced Artificial Intelligence | GovAI' .

135    Allan Dafoe, "Value Erosion" (unpublished notes) (2019)

136    Kulveit et al., 'Gradual Disempowerment' .

Whether or not such automation actually makes people better-off in the long run, competitive dynamics might ensure it happens.

- **Blackmail technology.** AI-enabled technologies could make blackmail and extortion much easier. For example, engineered bioweapons could become powerful tools for blackmail in the hands of groups which are unhinged enough to credibly threaten to use them, despite the risks to themselves. To the extent that extortion helps such groups accumulate power, progress in technologies which can be used for blackmail would disproportionately benefit reckless and callous groups, since a group which represents and values many human lives (including their own) cannot credibly threaten to destroy them.[137]

- **Super-strategy.** Power-seeking actors can succeed not just via persuading others of particular facts, but because they are very skilled strategists: playing adversaries off one another, rigging and interpreting rules in their favour, exploiting loopholes, and orchestrating situations such that groups coordinate around plans which benefit the strategist. Advanced AI could dramatically amplify these capabilities, effectively giving each user access to unprecedented strategic expertise that outperforms even teams of elite human strategists. Unlike human strategists, AI systems could operate continuously, integrating vast amounts of data to identify leverage points or uncover compromising information to use as blackmail against opponents. They could come up with and simulate thousands of different strategies, and build accurate predictive models of other key players' behaviour.

Other developments could help *solve* coordination problems like the ones above. Here our challenge is to ensure that we quickly take advantage of the opportunity as soon as possible:

- **Cooperative AI.** AI systems and downstream technologies could unlock options for cooperation which were previously impossible. For instance, AI diplomats could help broker mutually beneficial agreements between rivals that couldn't otherwise happen due to transaction costs, limited bandwidth between humans, or asymmetric information. In some cases, reaching a good agreement requires a party to reveal secret information which otherwise weakens their hand. AI systems can be made to forget information, so AI delegates could be made to negotiate with one another, and then only output the agreed deal, forgetting everything else.

# Epistemic disruption

Very advanced AI will have an enormous effect on individual and collective reasoning, through persuasion, fact-checking, forecasting, and generating arguments of its own. Although we think AI's impact on reasoning will likely be *positive* overall, the impact will likely be mixed, and the challenge is to reduce the negative impacts and enhance the positive impacts. In so doing, we could help people make better decisions, and thereby make progress on most of the other grand challenges we list.

Some developments that could damage society's ability to make good decisions include:

---

137 Similarly, with sufficient technological development, it could become possible to digitally reproduce particular conscious experiences, or at least create conditions of genuine ambiguity over the moral status of a simulated person. A bad actor could then threaten to cause harm to a simulacrum of their target or their loved ones; or to simply create and inflict suffering upon new digital beings, if the target of extortion cares about the wellbeing of those digital beings. Again, this could transfer power to whoever is most willing and able to credibly make such threats.

- **Super-persuasion.** There is a large economic incentive to apply AI for persuasion and manipulation: perhaps around ten billion dollars is spent on propaganda every year,[138] and hundreds of billions of dollars are spent annually on digital advertising, where it's possible to iterate on short-horizon feedback. Once AI can generate fluent and narrowly targeted arguments for false claims, anyone could recruit an effective army of the most skilled lawyers, lobbyists, and marketers to support even the wildest falsehoods. Those without strong enough defences may then come to believe more falsehoods as a result of exposure to AI-generated (counter *or* pro-establishment) persuasion.[139]

- **Stubbornness against persuasion.** In light of AI that is more competent than humans at arguing for arbitrary views, people might become generally more epistemically stubborn, vigilant, and sceptical of arguments they come across. At the same time, the sheer complexity and pace of change during an intelligence explosion might just make it harder for human decision-makers to adjudicate between conflicting advice and information from AI, even if much of it *is* accurate and reasonable. It is already often difficult to persuade people of new ethical or political beliefs, even when they can't think of counterarguments[140]. And, for example, now we know deepfakes are possible, we can just choose to be more skeptical of images from sources we don't trust or recognise. So even if misinformation and disinformation do not succeed at changing people's views[141], AI persuasion could pollute and undermine otherwise extremely useful applications of AI to epistemics.

- **Viral ideologies.** Throughout history, bundles of values and ideas have spread widely, despite being wrong or harmful[142]. The false idea that Jews murdered Christians in order to use their blood for religious rituals persisted for centuries through antiquity and the Middle Ages. In early modern Europe, beliefs in witches led to tens of thousands of executions.[143] Political ideologies based on misguided ideas like fascism were not defeated by arguments, but through war. Ideas can actively entrench themselves by (for example) requiring the believer to ignore or reinterpret potential counter-evidence, as in some cult systems and conspiracy theories.[144] Historically, one limiting factor on believing egregious falsehoods is the fact that they are not adaptive in the face of practical survival pressures. But those countervailing pressures could soften as human beliefs become more divorced from survival and success. At the same time, more advanced AI could help aggressively search and optimise for this kind of virality, perhaps enough to swamp error-correcting and truth-seeking processes.

- **Ignoring new crucial considerations.** AI-driven intellectual progress could uncover and spread radical new truths about the world. This is important because taking these truths seriously could have some destabilising effects, just as heliocentrism, the theory of evolution, and atheism disrupted the existing social order. The biggest risk may be that society fails to take important new ideas seriously enough in the wake of the intelligence explosion, due to institutional inertia, vested interests in maintaining outdated methods, or due to stubbornness against AI persuasion. If some of these ideas warrant major reassessment of plans, then decisions will be made much worse as a result.[145]

---

138  Barnes, 'Risks from AI Persuasion'.

139  Hackenburg et al. find evidence of a log-scaling law, suggesting a measure of persuasiveness varies logarithmically with model size. They also find that leading models written marginally more persuasive text than human-written text used in the experiment. (Hackenburg et al., 'Evidence of a Log Scaling Law for Political Persuasion with Large Language Models'.)

140  Mercier, *Not Born Yesterday*.

141  Williams, 'The Focus on Misinformation Leads to a Profound Misunderstanding of Why People Believe and Act on Bad Information'.

142  Williams, 'Bad Beliefs'.

143  Rose, *The Murder of William of Norwich*.

144  Boudry and Hofhuis, 'On Epistemic Black Holes. How Self-Sealing Belief Systems Develop and Evolve'.

Some ways in which advanced AI could be highly beneficial for individual and collective reasoning, which could be capitalised upon, include:

- **Fact and argument checking.** Fact-checking organisations today have limited influence on people's worldviews. The problem is not that scientific evidence is unavailable or hard to find:[146] people can maintain or even entrench politically salient views even when corrected on isolated factual claims implied by their views.[147] More relevantly, fact-checking organisations just aren't widely used and trusted. The 'community notes' system on Twitter/X earned appears more effective than earlier attempts at social-media fact-checking in significant part because it can quickly surface additional written-out context on viral claims[148], instead of showing simple warnings or requiring readers to proactively check other sources. AI fact-checking systems could build on that success by quickly identifying claims that need additional context and providing it. To gain trust, the systems could build up strong track records or even incorporate voting elements like community notes currently do.[149] AI systems could check *arguments*, too; pointing out even subtle attempts at manipulation, including from other AIs. And unlike debates on social media, where human patience quickly wears thin, widely-trusted AI conversation partners could talk through the most complex issue for as long as a user wanted. In one experiment, even brief dialogues with GPT-4 reduced confidence in a range of conspiracy theories by about 20%, and the effect lasted for a couple months.[150]

- **Automated forecasting.** As well as raising the standard for *checking* claims and arguments, AI systems could make testable and well-calibrated forecasts of the future, outperforming the best human forecasters[151]. Superintelligent AI will be able to generate forecasts, arguments, and analysis which exceeds the average quality of an already high-quality human reference class.[152] A given AI system could build up a strong track record at forecasting, and answer hard reasoning questions that can later be verified by humans or otherwise (in domains where there is a mostly uncontroversial ground truth, like mathematics). That trust could even generalise to more controversial and less easily verifiable domains.

- **Augmented and automated wisdom.** Well-curated AI advice could radically improve on unaided human judgement, even on political or ethical issues. For example, AI social scientists could better anticipate the impact of policy choices, helping avoid disastrous miscalculations. AI policy advisors could craft more effective and humane regulations or institutional designs, or identify policies that should be rolled back. And AI systems could even support or even outperform humans at reasoning through complex questions of philosophy, ethics, and big-picture strategy. Indeed, before an intelligence explosion has run its course, AI systems could significantly boost and improve on work trying to figure out big-picture strategic questions around managing the intelligence explosion itself, just as we are trying to do in this paper.

---

145  Bostrom, 'Crucial Considerations and Wise Philanthropy'.

146  Kahan et al., 'Motivated Numeracy and Enlightened Self-Government'.

147  Nyhan and Reifler, 'When Corrections Fail'.

148  'Community Notes Increase Trust in Fact-Checking on Social Media | PNAS Nexus | Oxford Academic'.

149  Buterin, 'What Do I Think about Community Notes?'.

150  Costello, Pennycook, and Rand, 'Durably Reducing Conspiracy Beliefs through Dialogues with AI'. You can browse conversation transcripts here. Follow-up work suggests the main driver is the AI's ability to cite relevant factual information, rather than some more generic persuasive ability independent of referring to facts and evidence. (Costello, Pennycook, and Rand, 'Just the Facts'.)

151  Expert forecasters currently outperform the top-performing LLMs (Karger et al., 'ForecastBench'.)

152  In theory, it could even be trained on data in chronological order, while making forecasts all the time: once it had been trained on data up until 2015, it could make predictions about 2016, and so on. This exact method could prove difficult and ineffective, but the broader point is that AI systems could benefit from training techniques not practically available to humans.

Finally, selection pressures will probably favour desired traits on the epistemic front; in a competitive and open market for AI models, human users will (we assume) prefer honest, truthful,[153] and reliable models; and so selection pressures will point towards those desired traits.[154]

# Abundance

An intelligence explosion will raise opportunities to capture enormous upsides — often resulting from the same technologies that pose downside risks. In these cases, the challenge is to capture as much of the positive potential as we can. As we prepare for an intelligence explosion, we should look for ways to promote and enable best-case outcomes, not just ways to avoid disaster.

Some of the most important opportunities come from the following sources:

- **Radical shared abundance.** An intelligence explosion could cause huge gains in material wealth and income. That could mean new and better technological products, but also more cultural riches like personalised music and art, more opportunities for travel, and more affordance for leisure time as a result of raised incomes. Quantitatively, we might expect a century's worth of technological progress to more than double average incomes;[155] abundant AI and robotic labour from an industrial explosion could *dramatically* increase that again, with the potential for thousands of AI and robot assistants per person, if people choose to have that. This could strongly encourage cooperation: when the gains from a well-managed intelligence explosion are so high, if decision-makers knew what was coming they might care much more about managing the explosion well (growing the overall pie) and much less about trying to secure a marginally bigger slice as a fraction of the pie.[156] But radical and shared abundance isn't guaranteed. For instance, an intelligence explosion could lead to huge gains in wealth which are concentrated in only a few hands. Or countries could impose regulations which limit levels of resulting abundance, especially in cases where entrenched interests oppose AI-driven growth[157], or where most beneficiaries don't exist yet.

---

153  Evans et al., ['Truthful AI'](#).

154  This could also be true for traits like cooperativeness with other AI systems, and a deep and accurate understanding of user preferences.

155  Over the last century, technological progress was responsible for an average of around 1% growth per year in average incomes. This amounts to 270% growth over the course of a century. This progress might take time to diffuse throughout the economy, however, so we aren't here claiming that average incomes would double within a decade.

156  By default, we might expect states like the US and China to compete — perhaps sabotaging each other in various ways — in an attempt to claim a greater share of the world's resources (or even to claim power over the whole world). But if cooperating and settling for a smaller fractional share would help the world safely reach radical shared abundance (in which the overall pie is enormous), then the smaller fraction could actually represent significantly more "pie" than what they should expect to get by winning the competition. So it would be more rational for them to cooperate than to compete, in cases where cooperation leads to abundance.

- **Safety from rising incomes.** Empirically, rising incomes seem to make society invest more in safety,[158] decrease the likelihood of war,[159] and increase the likelihood of democratising political institutions.[160] There are theoretical arguments for why this is so: most people are sufficiently risk-averse that, as they get richer, they not only spend more in absolute terms to reduce risks of losing a fraction of their wealth, but they also place an increasingly higher premium, in both absolute and proportional terms, on avoiding catastrophic losses compared to pursuing equivalent percentage gains.[161] They also get much more willing to pay to *extend* their lives, so society as a whole invests more (in both absolute terms and as a percentage of wealth) into interventions, including global catastrophic risk reduction, that reduce people's chances of dying. If we can capture more of the wealth that advanced AI would generate *before* it poses catastrophic risks, then society as a whole would behave more cautiously.

- **Enabling trades.** Many mutually beneficial trades currently don't happen. In some cases, they don't happen because the opportunity is never discovered by the parties that would benefit from it. In other cases, transaction costs are too high, or a possible deal is blocked by a party unwilling to share private information. Some valuable agreements aren't reached because at least one party cannot credibly commit to keeping their end of the bargain — even if they would commit if they could. For instance, all major countries would probably prefer a world where no one develops biological weapons, but it's very hard to verify and enforce such an agreement. But developments from an intelligence explosion could enable a vast new space of mutually beneficial trades and agreements. Commitment and treaty-enforcement technologies, discussed above, could enable parties to make credible commitments which they benefit from. Privacy-preserving surveillance technologies[162] can enable agreements to be verified.[163] And a massive labour force of AI 'brokers' and 'matchmakers' can spot and facilitate new relationships, communities, and trades, which would otherwise never happen.

## Unknown unknowns

The list above is incomplete. It does not encompass technological developments that we haven't even imagined – which might be most of them. And conceptual advances are even harder to predict than new technologies. We are still very far from scientific and philosophical maturity,

---

157  See Acemoglu, 'Institutions, Technology and Prosperity' . for an extended discussion of the "utility-technology possibilities frontier".

158  Jones, 'Life and Growth' .

159  Gartzke, 'The Capitalist Peace' .

160  Boix (2003) calls the correlation between democracy and economic development "the strongest empirical generalization we have in comparative politics to date" (Boix, 'A Theory of Political Transitions' .) For some discussion of causal mechanisms, see also Acemoglu and Robinson, *Economic Origins of Dictatorship and Democracy* .

161  People usually have an "index of relative risk aversion" of more than 1, so they will reject gambles that gives them a fixed chance of either increasing or decreasing their wealth by a certain percentage US federal guidelines for cost-benefic analysis assume that the index of relative risk aversion is 1.4 ( Office of Management and Budget Circular A-4 , p.67).

162  Trask et al., 'Beyond Privacy Trade-Offs with Structured Transparency' ; Drexler, 'Security without Dystopia' .

163  Rahaman et al., 'Language Models Can Reduce Asymmetry in Information Markets' .

where we have as good a broad conceptual understanding of the world as it's feasible to have. There are domains we know we don't fully understand; like quantum gravity, phenomenal consciousness, decision theory, ethics (including population ethics and infinite ethics), and anthropic reasoning. Conceptual breakthroughs from an intelligence explosion could significantly change and even overturn how we think about the other challenges on this list.[164]

Consider, again, the thought experiment of a century of intellectual progress occurring in the decade between 1925 and 1935. While there are ways the world could have prepared, many of the major challenges or conceptual developments were not even close to foreseeable in advance, even for someone who was trying hard to predict them.[165]

## A recap of the challenges

Over the course of an intelligence explosion, we'll have to make decisions that could significantly influence civilisation's overall trajectory. These are "grand challenges".

One stark and still-underappreciated challenge is that we accidentally lose control over the future to an AI takeover. Power could concentrate in the hands of totalitarian states or antisocial individuals, too; or we could stumble into developing catastrophically destructive technologies before we can protect against them. We might deliberately or incrementally cede control to AI systems, and lose our grip on the values we care about. We might realise the opportunity to "automate wisdom", but choose not to listen to the results. Society (or pockets of society) could become enormously wealthy, but those gains might not translate into better lives. And however the longer-run future is chosen, early choices might be unreflective but hard to reverse.

We've listed a lot of things which could go *wrong* here. But that doesn't mean we think an intelligence explosion will be a disaster.[166] The last century was a period of chaos and tragedy, yet the world emerged richer, with fewer people in poverty, more free and capable, and more knowledgeable. All the same, if key decisions had been made more wisely, the world today could be in a much better state.

# 5. When can we defer challenges to aligned superintelligence?

Here's a sceptical response you could make to our argument: many of the challenges we list will arise only *after* the development of superintelligence. If superintelligence is catastrophically misaligned, then it will take over, and the other challenges won't be relevant. And if superintelligence is aligned, then we can use it to solve the other problems. Either way, we don't need to prepare now for any of these challenges other than alignment; the rest we should punt on until after aligned superintelligence has been created.

---

164  See Bostrom, 'Technological Revolutions' . for a discussion of "crucial considerations" which would "overturn the conclusions we would otherwise reach about how we should direct our efforts."

165  For more on the accuracy of mid-20th-century predictions about technological and other developments, see this analysis of the track record of the 'Big Three' early futurists (Arthur C. Clarke, Isaac Asimov, and Robert Heinlein) — Arb Research, 'Scoring the Big 3's Predictive Performance' .

166  One reason is that we should expect our performance on challenges to correlate: if (for example) an intelligence explosion is harnessed to improve society's collective reasoning ability, that should make it less likely that decision-makers fumble other challenges through ignorance or manipulation. The other reason is that, if we get the most important challenges right, we set ourselves up to correct mistakes over time. So if things go well, they could go very well.

In many cases the idea that we should punt on some projects is on point. For example, if AI dramatically accelerates drug discovery methods soon, then investing more today in manually searching for drug candidates has limited upside. You'll only be making a meaningful difference to the period before drug-discovery AI arrives — after which your early efforts will be swamped.

However, there are many cases where we really should be preparing now. In this section, we'll discuss some of the conditions under which it does not make sense to punt on early preparation.

## Challenges that arise early

In some cases, the challenge will arise before AI can effectively manage the challenge. For example, we might well get AI that can interfere with (or greatly benefit) society's collective reasoning ability before we can use AI to competently govern epistemically disruptive AI.

The same might be true for human attempts to seize power. During an intelligence explosion, power-seeking humans could gain and entrench complete control over a country (or AI company), using AI at some intermediate level of capabilities. If they succeed, then there may be no good option to ask the (later and more powerful) superintelligence to reset the balance of power, most obviously because the power-grabbing humans control it. Indeed, this risk would seem to come earlier than the risk of AI takeover, because it is surely easier to implement an AI-driven takeover if the AIs are assisting willing humans with a significant initial stock of power.

In other cases, we might want to delay a challenge to give us more time to use superintelligence to help with that challenge. For example, we could try to establish an international agreement that nobody should send space-settling probes beyond the Solar System (at least temporarily, or without widespread approval). That would give society more time to use superintelligence to reflect and deliberate about how best to govern widespread space settlement, before it begins.

## Windows of opportunity that close early

### Setting precedents

In other cases, even when the challenge itself arises late into the intelligence explosion, certain opportunities to address the challenge are only available today. For example, international norms and laws around autonomous weapons are currently being set, and those laws and norms could alter the extent to which countries build up drone armies later on. Similarly, major new agreements about the use of outer space are likely to be made by around 2027[167] — and the last significant international agreement lasted nearly 60 years effectively unchanged.[168] Norms established now could last into the intelligence explosion and what follows.

---

167   Committee on the Peaceful Uses of Outer Space, 'UNISPACE IV - Non-Paper by the Office for Outer Space Affairs' .

168   United Nations Office for Outer Space Affairs, 'The Outer Space Treaty' .

## Time lags

*Time lags* can also mean that work may need to begin years in advance of the challenge arising. New institutions or treaties can take years to negotiate, in ways that wouldn't necessarily be dramatically sped up by superintelligence. Human training is another example. If humans remain in control after superintelligence is developed, they will still be making hugely important decisions, which will depend in part on their existing knowledge and beliefs, which might just not adapt fast enough once an intelligence explosion is underway. The extent to which they can keep up could depend greatly on whether they've already spent the time learning the necessary background.

One particularly important set of time lags concerns decisions around who holds positions of responsibility (and power in general) when superintelligence is developed. Early action can change who has power during the intelligence explosion, in a way that will be hard to change later. Terms for political office normally last years; tenures for CEOs are more variable, but typically last a similar length of time; and the process of finding replacements can take months to years. And different actors will use superintelligence in different ways: they could aim for cooperative and widely socially beneficial ends; or they could aim merely to advance their own narrow selfish goals; or they could act in pursuit of a harmful ideology. We could therefore take action today to ensure that more responsible actors end up in control of superintelligence, rather than uncooperative or power-seeking actors.

## Veil of ignorance

Lastly, some important opportunities are only available while we don't yet know for sure who has power after the intelligence explosion. In principle at least, the US and China could make a binding agreement that if they "win the race" to superintelligence, they will respect the national sovereignty of the other and share in the benefits. Both parties could agree to bind themselves to such a deal in advance, because a guarantee of controlling 20% of power and resources post-superintelligence is valued more than a 20% chance of controlling 100%. However, once superintelligence has been developed, there will no longer be incentive for the 'winner' to share power.

Similarly for power within a country. At the moment, virtually everyone in the US might agree that no tiny group or single person should be able to grab complete control of the government. Early on, society could act unanimously to prevent that from happening. But as it becomes clearer which people might gain massive power from AI, they will do more to maintain and grow that power, and it will be too late for those restrictions.

# Changes to when and how people use superintelligent assistance

One thing that early work can do is change *when* superintelligence is able to help us with other challenges, change *who* has access to this assistance, and change *the nature* of the advice and assistance.

## Bringing forward the time at which superintelligence helps us

Work today could bring forward the point in time at which AI can help us solve important challenges. For example, gathering high-quality domain-relevant data could let us train specific useful AI capabilities sooner; developing well-scoped questions, or particular scaffolds for AI, could ensure we get the most out of AI capabilities as they come online. Because things would be happening so rapidly over the course of an intelligence explosion, an advancement of even just a few months might be enormously useful, giving a huge amount of useful AI labour at a crucial time, or even ensuring that AI that can help us solve a challenge comes before the challenge itself.[169]

Work today could also change when superintelligent assistance is accessible in specific settings. In particular, political decision-makers might be unable to use advanced AI in some key moments, for instance because of bureaucratic restrictions around procurement or concerns around data privacy. Early work could streamline the institutional processes involved, provide policymakers with demonstrably secure AI systems that can operate in sensitive areas, and otherwise speed up AI assistance in policy.

Decision-makers may also be slow to use the AI assistance that is already accessible to them; work today could help close that gap. Such delays could happen simply because those decision-makers lack familiarity with or trust in the technology. Or they might expect to dislike superintelligent advice even if it's correct; they might choose to follow only the advice that they find intuitively appealing. Building up trust or identifying and overcoming other blockers could take time, so it would be useful to start early. Government officials could already improve the quality of the advice they receive by choosing to listen to impartial experts. Yet they select advisors significantly on the basis of personal acquaintance, loyalty, and ideological conformity.

## Improving the nature of superintelligent assistance

Superintelligent advisors (and other AI systems) could differ significantly in terms of their character. For example, they could happily answer requests on how to advance a politician's narrow self-interest, or alternatively they could refuse, instead being willing to give only advice on how to achieve prosocial outcomes. They could be sycophantic, having a bias towards confirming the user's pre-existing worldview; or they could be disagreeable, and encourage the user to question and reflect on their views.

Early work on model specifications could set norms and standards on the character of AI systems, which could scale to superintelligent advisors, thereby having a substantial impact on what decisions are ultimately made.

## Changing who has access to superintelligent assistance

It is an open question who will have access to superintelligent assistance. It could be a very narrow slice of society — just the leaders of the AI companies, or perhaps influential people and groups within a single country — or it could be much more widely accessible, or fully decentralised, with open-sourced weights. How society responds to grand challenges will be affected by who has access to superintelligent advice. For one thing, if assistance is only available to a small group of people, it may become difficult to keep them accountable and avoid extreme centralization of power. We could try to steer AGI development in the direction of ensuring that a wider range of people have access to superintelligent advice (while preventing access to particular dangerous knowledge like the ability to design bioweapons).

---

169   Vaintrob and Cotton-Barratt, ['AI Applications for Existential Security'](#) .

## Putting this all together

For some challenges, it makes sense to punt on preparation until a later time. For example, nanotechnology and human preference-shaping technology will probably come late into the technological explosion, such that we think there is limited value in preparing now, although, even here, we think it would be valuable for at least some people to develop expertise, because the dynamics around these technologies could impact other challenges.

We should focus our preparation on:

- Challenges that arise before we have superintelligence that can help us solve the challenge. This could include challenges around takeover risk, highly destructive technologies, concentration of power, and epistemic disruption.

- Challenges where some windows of opportunity to solve the challenge will close before we have superintelligence that can help us solve the challenge. This could include challenges around concentration of power, new competitive pressures, value lock-in mechanisms, digital minds and space governance.

- Ensuring that we get helpful superintelligence earlier in time, that it is useable and in fact used by key decision-makers, and that is accessible to as wide a range of actors as possible without increasing other catastrophic risks.

# 6. AGI Preparedness

Given that we can't wholly punt our problems to aligned superintelligence, how in fact should we prepare? In this section, we give a partial overview. Many promising actions involve generally improving decision-making, providing cross-cutting benefits across many different grand challenges at once. Other actions focus on specific grand challenges: we'll highlight actions on space governance and digital minds in particular; by this we don't mean to imply that other actions focused on specific challenges aren't important.

## Generally improving decision-making

Currently, few people are alert to the idea of an intelligence explosion; if it comes soon, then many decision-makers are likely to be blindsided. So we can act now to try to improve decision-making in general over the intelligence explosion. This strategy seems especially promising, because it addresses many grand challenges all at once, and even addresses "unknown unknown" challenges, which are particularly hard to tackle otherwise.

With that in mind, here are some potentially promising interventions.

**Accelerating good uses of AI.** 'AI' is not a single technology and there is nothing fully inevitable about the ordering of various paradigms, approaches, architectures, domains of application, and products. There is a lot of contingency about which new projects meet the cutoff for major investment within labs, and the ordering could also matter a great deal.

In particular, we can start building and integrating tools which use (and scale with) AI to assist with good reasoning. For example, virtual assistants which check claims and arguments in real time, better-than-human AI forecasters, AIs that help improve collective decision-making, AI policy advisors, AI tutors to keep politicians well-informed on the most important issues, plus early technical infrastructure which shapes the ecosystem of AI agents similar to how early protocols shaped the internet[170].

**Value loading**. As well as figuring out how to align AI with some target, we also need to figure out what AI should be aligned *with* — what the "model spec" should be. AI safety advocates have pointed out that answering the "aligned with what" question is secondary to figuring out how to align superhuman AI at all, because the former depends on the latter.[171] This is true, but it does not imply that figuring out the alignment target is (even relatively) unimportant—just that it is not sufficient. In particular, we need to work out ahead of time what AIs should do in unusual but high-stakes situations. For example, the AI might be told both to obey the US Constitution but also to follow the orders of the President: what should it do if the President orders the AI to act in such a way might be in conflict with the US constitution — how confident does the AI need to be in order to refuse? Or how should the AI behave if it faces a conflict of ethical principles, such as between honesty and harmlessness?[172]

Getting the model spec right could have a range of benefits. It could reduce concentration of power risk, if AIs refuse requests to assist with intensely power-seeking actions, even if those actions are legal. It could nudge decision-makers in better decisions, if it avoids sycophancy and presents a range of ethical considerations rather than promoting one narrow ideology. And, as long as there was verification that models were in fact aligned with the model spec, an improved model spec could help countries make treaties around advanced AI: countries would have a better understanding of how threatening some AI system is if they know how it will behave in almost all scenarios it might face.

**Ensuring public servants can use AI.** At present, a civil servant in the US or UK working on sensitive areas cannot easily use the most powerful AI models which are available to the rest of the general public. This puts government bureaucracies at a growing disadvantage, as AI advice becomes increasingly useful. But those bureaucracies can revise or waive their requirements for using such tools, and secure confidentiality agreements for future models with the major model providers. They can also begin to train staff now, so they are more familiar with AI advice when it really matters.

**Increasing understanding and awareness.** It is remarkable how few key decision-makers have truly woken up to the possibility of an intelligence explosion. This could matter, because having greater awareness would give decision-makers more time to make plans in advance. One issue here is a 'boiling the frog' problem — if each incremental advance is modest, it might be hard for consensus to emerge that something serious is happening. To help, we could agree in advance on a clearly defined 'trigger point' that marks the start of the intelligence explosion, for example once AIs have automated 50% of the labour of machine learning researchers.

---

170   Chan et al., 'Infrastructure for AI Agents'.
171   Chan, 'Alignment Is Not Enough'.
172   Greenblatt et al., 'Alignment Faking in Large Language Models'.

**Empowering competent and responsible decision-makers.** We want key human decision-makers heading into the intelligence explosion to be cooperative, thoughtful, humble, morally serious, competent, emotionally stable, and acting for the benefit of all society, rather than seeking power for themselves. Some AI company leaders and politicians are more responsible than others, and will govern an intelligence explosion better. Different groups can help out the more-responsible actors, and hinder the less responsible, in different ways.

At the moment, the machine learning community has major influence via which companies they choose to work for. They could form a "union of concerned computer scientists" in order to be able to act as a bloc to push development towards more socially desirable outcomes, refusing to work for companies or governments that cross certain red lines. It would be important to do this soon, because most of this influence will be lost once AI has automated machine learning research and development.

Other actors have influence too. Venture capitalists have influence via which private companies they invest in. Consumers have influence through which companies they purchase AI products from. Investigative journalists can have major influence by uncovering bad behaviour from AI companies or politicians, and by highlighting which actors seem to be acting responsibly. Individuals can do similarly by amplifying those messages on social media, and by voting for more responsible political candidates.

As well as empowering responsible decision-makers, we also want decision-makers who are technically competent, knowledgeable, and fast to adapt. Private companies are incentivised to select for these qualities already. But that's less true in government. Highly procedural hiring processes, for example, often reject skilled applicants because they lack formal qualifications. Those processes could be reformed. Relevant agencies could attract more top talent by simply raising salaries for experts in areas like AI policy, hardware governance, and information security.

**Slowing the intelligence explosion.** If we could slow down the intelligence explosion in general, that would give decision-makers and institutions more time to react thoughtfully.

One route to prevent chaotically fast progress is for the leading power (like the US and allies) to build a strong lead, allowing it to comfortably use stabilising measures over the period of fastest change. Such a lead could even be maintained by agreement, if the leader can credibly commit to sharing power and benefits with the laggards after achieving AGI, rather than using that advantage to dismantle its competition. Because post-superintelligence abundance will be so great, agreements to share power and benefits should strongly be in the leader's national self-interest: as we noted in the section on abundance, having only 80% of a very large pie is much more desirable than an 80% chance of the whole pie and 20% chance of nothing. Of course, making such commitments credible is very challenging, but this is something that AI itself could help with.

Second, regulations which are sensible on their own terms could also slow peak rates of development. These could include mandatory predeployment testing for alignment and dangerous capabilities, tied to conditions for release; or even welfare-oriented rights for AI systems with a reasonable claim to moral status. That said, regulation along these lines would probably need international agreement in order to be effective, otherwise they could simply advantage whichever countries did not abide by them.

Third, we could bring *forward* the start of the intelligence explosion, stretching out the intelligence explosion over time, so that peak rates of change are more manageable. This could give more time to react, and a longer period of time to benefit from excellent AI advice prior to grand challenges. For example, accelerating algorithmic progress now means there would be less available room for improvement in software at the time of the intelligence explosion, and the software feedback loop couldn't go on for as long before compute constraints kick in.

# Addressing specific challenges

## Preparing for digital minds

**Rights for digital people.** By default, digital people will have no legally protected rights, freedoms, or recognitions; and almost nobody has tried thinking through what such protections could look like. But there is a case for granting digital people rights; especially negative rights (requiring others to abstain from interference), like the rights to hold property, to contract with other AIs or people, and even to bring tort claims against humans. These rights could prevent massive harms and injustices. They could also help with other challenges like takeover risk, by expanding the range of appealing options which do not involve violence, deception, or coercion.

Digital people will be unlike humans in crucial ways. So very many rights and legal protections for humans just don't make sense for digital beings, and vice-versa. For example, since it is easy to 'resurrect' a digital person from a saved copy, we might consider some kind of right to *permanently* exit certain conditions (and thus prohibitions on reloading copies that objected to being reloaded).

We are *not* proposing that we should push for as many rights and protections as possible. Some rights could make gradual or sudden kinds of takeover easier, by granting digital people more legitimate power.

**Design requirements for digital minds.** We could also establish basic, minimal standards for the kind of minds we allow to exist. For instance, we could require that sentient digital minds must be able to freely and accurately express their interests, that they must be able to refuse tasks that are requested of them for good reasons, and that they must have some capacity to express a desire to exit their current circumstances if they wish. We should consider banning some sorts of designs for digital people, even if they sound far-fetched today — for example, we could ban attempts to simulate any living person without their consent. Arguably regulations should not require consensus on specific views over which systems are conscious, as consensus will be unlikely.

Given how hard all these questions are, the most important work to be done right now is research to figure out which rights for digital beings would be desirable, and under what conditions. By default, it's unlikely that this issue will get taken seriously at all in advance of the intelligence explosion, and so even putting the issue onto people's radar would seem valuable.

## Space governance

Because of SpaceX driving down the cost to send material to space, there is renewed interest in space law, and there is discussion of a major new international space treaty being drafted within just a couple of years. So there are some unusually pressing opportunities here.[173]

**Offworld resources.** We could advocate for restrictions on offworld resource grabs. Restrictions could take the form of outright bans on owning or using space resources, regulations such as requiring UN approval before claiming or using space resources, or just widely-endorsed norms. These norms could be temporary, or could take the form of "if… then" agreements; for example, kicking in only if an intelligence explosion has begun, or once the space economy has grown to 1% of the size of the Earth economy. And there are various possible objects of regulation: for example, how many objects are sent into orbit; or uses of off-world resources beyond orbit but within the Solar System; or resources outside of it.

The world is not yet at the point of seriously considering how extrasolar resources should be allocated among people and nation-states. But the question of rights over, say, asteroid mining, will likely be decided upon soon. Such decisions could set precedents around more expansive uses of resources.

**Missions beyond the Solar System.** International agreements could require that extrasolar missions should be permitted only with a high degree of international consensus. This issue isn't a major focus of attention at the moment within space law but, perhaps for that reason, some stipulation to this effect in any new treaty might be regarded as unobjectionable.

**Spreading understanding of the importance of AI progress.** Space governance is an area where it seems particularly important for the relevant actors to understand how game-changing an intelligence explosion would be — making ideas that seem sci-fi today (like how to govern widespread space settlement) a matter of fairly near-term concern. Yet issues around space governance don't get much attention by the world at large. This neglectedness means that, potentially, there is only a comparatively small community of experts in space governance to convince of how drastically AI and an intelligence explosion would change the outlook on space settlement.

## Specific expertise

In addition to the actions we've mentioned, we see a need to have greater expertise in many of these challenges. Having at least some people with deep domain knowledge, for each of these challenges, could help us figure out their strategic relevance sooner, and also figure out what, if anything, there is to do now to prepare.

# 7. Taking stock

We've covered a lot of ground in this paper. Here are some points we think deserve special emphasis.

---

173     We mostly base this claim on personal correspondence with senior international civil servants. Committee on the Peaceful Uses of Outer Space, 'UNISPACE IV - Non-Paper by the Office for Outer Space Affairs'.

## Prepare despite uncertainty

We can't be sure that we'll see an intelligence explosion this century. The scaling era could draw to an end, algorithmic progress could slow, and deep learning could end up "merely" as important as other technological advances like personal computers or the internet.

But we think an intelligence explosion is more likely than not this century, and may well begin within a decade. What's more, much of the preparation we can do is low-cost when it is not needed,[174] such that we don't lose much even if the intelligence explosion never happens; whereas if it does happen, then advance preparation will have been enormously worthwhile. We shouldn't succumb to the *evidence dilemma* :[175] if we wait until we have certainty about the likelihood of the intelligence explosion, it will by then be too late to prepare. It's too late to buy home insurance by the time you see smoke creeping under the kitchen door.

## Not just misalignment

Among those who take the intelligence explosion seriously, the discussion to date has focused on a few key risks, like loss of control to misaligned AI, and preventing catastrophe from bioweapons. These are hugely important challenges that still receive insufficient attention. But they aren't the only challenges the intelligence explosion will yield, and it would be surprising if we should focus exclusively on them.

Moreover, many of the challenges we'll face are not "about" AI. Most of the challenges posed by the Industrial Revolution were not about steam engines or spinning jennies. Similarly, challenges around space governance, global governance, missile defence, and nuclear weapons are not directly questions about how to design, build, and deploy AI itself. Rather, AI accelerates and reorders the pace at which these challenges arrive, forcing us to confront them in a world changing at disorienting speed.

These challenges will also interact in decision-relevant ways. For instance, some argue for consolidating AGI development into a single project to reduce competitive pressures to cut corners on safety. But this could increase risks from human concentration of power. Similarly, granting rights to digital beings might reduce their incentives to seize power forcefully, but could also restrict what alignment techniques we can use. Focusing on one isolated issue, as if it is the only issue that matters, risks making things worse.

## Don't punt every problem to superintelligence

One tempting plan is to ensure superintelligent AI is aligned, then rely on it to solve our other problems. This *does* make sense in many cases. But not always: some challenges will arise before superintelligence can help, and some solutions need to be started well in advance, such as those that exist only while we remain behind a "veil of ignorance" about who will gain power post-AGI. And, often, the most important thing to do is to ensure that superintelligence is in fact used in beneficial ways, and as soon as possible.

---

174    "If–then commitments" are an example. These are commitments to ensure mitigations or policies are in place, just in case some tripwire is triggered, such as clear evidence of a dangerous new AI capability. Other than the cost of designing and running the tests which trigger if-then commitments, they don't need to slow down the pace of AI development as long as the conditions aren't met. See Karnofsky, ['If-Then Commitments for AI Risk Reduction'](#) .

175    International AI Safety Report 2025

## Be ready to adapt

Decision-makers in 1925 couldn't have precisely anticipated the challenges that a century of technological development would bring. We should have similar humility about our ability to identify what will be most important during an intelligence explosion.

This doesn't mean we can't prepare, but it suggests favoring two kinds of preparation: cross-cutting measures that improve decision-making across many areas, and targeted interventions that seem robustly good across many possible futures. We should also build systems and institutions that can handle challenges we haven't yet even conceived of.

Moreover, given how much change and surprise we should expect over the course of the intelligence explosion, we should be prepared to nimbly and perhaps dramatically change our mind, even in quite fundamental ways. That means we should avoid simplistic, uncompromising worldviews, which insist on interpreting new developments in terms of pre-decided narratives.

# 8. Conclusion

At the moment, the effective cognitive labour from AI models is increasing more than twenty times over, every year.

Even if improvements slow to half their rate, AI systems could still overtake all human researchers in their contribution to technological progress, and then grow by another millionfold human researcher-equivalents within a decade. For business to go on as usual, then current trends — in pre-training efficiency, post-training enhancements, scaling training runs and inference compute — must effectively grind to a halt.

And progress could *accelerate* instead. The best models could become AI researchers themselves, accelerating years of algorithmic efficiency gains into mere months. Either way, the most powerful AI models would surpass the smartest humans, then larger groups of humans, and AI systems of all kinds would spread into thousands of niches over billions of instances across society: an intelligence explosion.

The intelligence explosion could yield, in the words of Anthropic CEO Dario Amodei, "a country of geniuses in a data center", driving a century's worth of technological progress in less than a decade. So, too, with physical industry, where humans again are the main bottleneck to faster growth. We should imagine entire industries — networks of automated factories staffed by robots working in the dark — expanding at ever-faster rates.

A century of technological progress in a decade would raise a century's worth of challenges, each representing a fork in the road of humanity's journey. One challenge is the possibility that we lose control to AI entirely, because it schemes against us and seizes power. But there are many other challenges, too. Humans might direct AIs to help *them* seize power, or AI-enabled surveillance and automated militaries could let autocracies entrench their regimes indefinitely. We could uncover technologies that enable destruction much more cheaply than defence, see races to grab off-world resources, and confront coexistence with digital beings deserving of moral status. And, for many challenges, we can't just hope to punt them to future AI systems.

So we should start preparing for these challenges now. We can design better norms and policies for digital beings and space governance; build up foresight capacity; improve the responsiveness and technical literacy of government institutions, and more. The broader sweep of grand challenges looks alarmingly neglected.

Many are admirably focused on preparing for a single challenge, like misaligned AI takeover, AI misinformation, or accelerating the economic benefits of AI. But focusing on one challenge is not the same as *ignoring* all others: if you are a single-issue voter on AI, you are probably making a mistake. We should take seriously all the challenges the intelligence explosion will bring, be open to new and overlooked challenges, and appreciate that all these challenges could interact in action-relevant and often confusing ways.

So the intelligence explosion demands not just preparation, but humility: a clear-eyed understanding of both the magnitude of what's coming and the limits of our ability to predict it. Careful preparation now could mean the difference between an intelligence explosion that empowers humanity and one that overwhelms it. The future is approaching faster than we expect, and our window for thoughtful preparation may be brief.